\documentclass[aps,pra,reprint]{revtex4-2}
\usepackage{graphicx, amsmath, amssymb, hyperref}
\usepackage[caption=false]{subfig}

\begin{document}


\newcommand{\comb}[2]{{\begin{pmatrix} #1 \\ #2 \end{pmatrix}}}
\newcommand{\braket}[2]{{\left\langle #1 \middle| #2 \right\rangle}}
\newcommand{\bra}[1]{{\left\langle #1 \right|}}
\newcommand{\ket}[1]{{\left| #1 \right\rangle}}
\newcommand{\ketbra}[2]{{\left| #1 \middle\rangle \middle \langle #2 \right|}}

\newcommand{\fref}[1]{Fig.~\ref{#1}}
\newcommand{\Fref}[1]{Figure~\ref{#1}}
\newcommand{\sref}[1]{Sec.~\ref{#1}}
\newcommand{\tref}[1]{Table~\ref{#1}}


\title{Quantum Search with a Generalized Laplacian}

\author{Jonas Duda}
	\email{jonasduda@creighton.edu}
	\affiliation{Department of Physics, Creighton University, 2500 California Plaza, Omaha, NE 68178}
\author{Molly E. McLaughlin}
	\email{mollymclaughlin@creighton.edu}
	\affiliation{Department of Physics, Creighton University, 2500 California Plaza, Omaha, NE 68178}
\author{Thomas G. Wong}
	\email{thomaswong@creighton.edu}
	\affiliation{Department of Physics, Creighton University, 2500 California Plaza, Omaha, NE 68178}

\begin{abstract}
	A single excitation in a quantum spin network described by the Heisenberg model can effect a variety of continuous-time quantum walks on unweighted graphs, including those governed by the discrete Laplacian, adjacency matrix, and signless Laplacian. In this paper, we show that the Heisenberg model can effect these three quantum walks on signed weighted graphs, as well as a generalized Laplacian equal to the discrete Laplacian plus a real-valued multiple of the degree matrix, for which the standard Laplacian, adjacency matrix, and signless Laplacian are special cases. We explore the algorithmic consequence of this generalized Laplacian quantum walk when searching a weighted barbell graph consisting of two equal-sized, unweighted cliques connected by a single signed weighted edge or bridge, with the search oracle constituting an external magnetic field in the spin network. We prove that there are two weights for the bridge (which could both be positive, both negative, or one of each, depending on the multiple of the degree matrix) that allow amplitude to cross from one clique to the other---except for the standard and signless Laplacians that respectively only have one negative or positive weight bridge---boosting the success probability from 0.5 to 0.820 or 0.843 for each weight. Moreover, one of the weights leads to a two-stage algorithm that further boosts the success probability to 0.996.
\end{abstract}

\maketitle


\section{Introduction}

The continuous-time quantum walk was introduced by Farhi and Gutmann \cite{FG1998b} as a technique for solving decision trees. In this analog model of quantum computing \cite{Childs2009}, the $N$ vertices of a graph label computational basis states $\ket{1}, \ket{2}, \dots, \ket{N}$, and the state is generally a superposition over the vertices. It evolves by Schr\"odinger's equation, and the kinetic energy of the system is captured in the Hamiltonian by the term
\[ H_L = -\gamma L, \]
where $\gamma$ is the jumping rate of the quantum walk, and
\[ L = A - D \]
is the discrete Laplacian (the discrete analogue of $\nabla^2$), where $A$ is the adjacency matrix ($A_{ij}$ is the weight of the edge joining vertices $i$ and $j$, which is zero if the vertices are nonadjacent) and $D$ is the degree matrix of the graph (a diagonal matrix where $D_{ii}$ equals the degree of vertex $i$, which is the sum of the weights of the edges incident to the vertex). Such \emph{Laplacian quantum walks} have also been utilized for quantum search \cite{CG2004} and state transfer \cite{Alvir2016}.

Another kind of quantum walk is obtained by dropping the degree matrix $D$, so the quantum walk is governed by the adjacency matrix $A$ alone, and its Hamiltonian is
\[ H_A = -\gamma A. \]
Such \emph{adjacency quantum walks} have also been used for quantum search \cite{Novo2015} and state transfer \cite{Godsil2012}, but they have also been used to solve boolean formulas \cite{FGG2008} and can solve a problem exponentially faster than classical computers \cite{Childs2003}.

A third kind of quantum walk utilizes the signless Laplacian,
\[ Q = A + D, \]
which differs from the standard Laplacian by the degree matrix taking a positive sign. Then, the Hamiltonian is
\[ H_Q = -\gamma Q. \]
Such \emph{signless Laplacian quantum walks} have also been employed for state transfer \cite{Alvir2016}, and recently, they were shown to solve a quantum search problem faster than the Laplacian and adjacency quantum walks, for certain parameter regimes of the complete bipartite graph \cite{Wong43}.

It has been shown that a single excitation in the Heisenberg (XYZ) model of a spin-network, with appropriate coupling constants, can implement these three quantum walks on \emph{unweighted} graphs \cite{Bose2009,Wong43}. In this paper, we show that the Heisenberg model can also implement these three quantum walks on \emph{signed weighted} graphs, and furthermore, they can implement a quantum walk governed by a generalized Laplacian
\begin{equation}
    \label{eq:Lalpha}
     L_\alpha = L + \alpha D,
\end{equation}
where $\alpha$ is a real-valued parameter, so it can be positive, negative, or zero and need not be an integer. Its Hamiltonian is
\[ H_\alpha = -\gamma L_\alpha, \]
and we call this a \emph{generalized Laplacian quantum walk}. The generalized Laplacian is motivated by the observation that
\begin{align*}
    L &= L + 0D = L_0 \\
    A &= L + 1D = L_1, \\
    Q &= L + 2D = L_2,
\end{align*}
and adding other multiples of the degree matrix to the Laplacian may be of interest, as could subtracting multiples or permitting non-integer multiples. In \sref{sec:spin-networks}, we will prove that the Heisenberg model, with appropriate coupling constants, implements $L_\alpha$ on signed weighted graphs.

\begin{figure}
\begin{center}
    \includegraphics{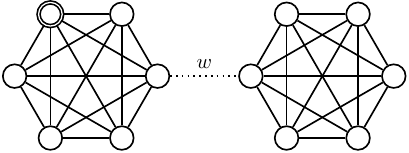}
    \caption{\label{fig:barbell}A weighted barbell graph of $N = 12$ vertices. Solid edges are unweighted, and the dotted edge has real-valued weight $w$. A vertex is marked, indicated by a double circle.}
\end{center}
\end{figure}

In \sref{sec:barbell}, we will explore an algorithm using the generalized Laplacian quantum walk. Specifically, we will explore spatial search, where an external magnetic field implements an oracle, on a weighted barbell graph, which consists of two equal-sized unweighted cliques connected by a single edge or bridge of weight real-valued $w$. An example is shown in \fref{fig:barbell}. Since $w$ can be negative, this is a signed weighted graph. This contrasts with \cite{Wong7} and \cite{Wong41}, which also explored search on the barbell graph, but respectively with an unweighted bridge ($w = 1$) only or with increasing weights ($w \ge 1$) only, and both with only the Laplacian ($L_0 = L$) and adjacency ($L_1 = A$) quantum walks. As we will prove, with real-valued weights and the generalized Laplacian, the weighted bridge is typically too restrictive for amplitude for move from one clique to the other, but there are usually two ``critical'' weights that do allow probability to cross the bridge, boosting the success probability of the search. The exceptions to this are the standard Laplacian and signless Laplacian, which each only have one critical weight. In contrast, in \cite{Wong41}, since only positive weights were considered, no critical weights were found for the Laplacian quantum walk, only one of two critical weights was found for the adjacency quantum walk. In addition, for the generalized Laplacian, one of the critical weights leads to a two-stage algorithm that boosts the success probability further. Thus, regardless of $\alpha$ \eqref{eq:Lalpha}, the weight $w$ can be chosen to improve the success probability over the unweighted $w = 1$ case. We conclude in \sref{sec:conclusion}.


\section{\label{sec:spin-networks}Spin Networks and Signed Weighted Graphs}

\begin{figure}
\begin{center}
    \includegraphics{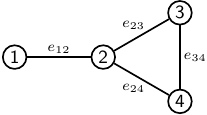}
    \caption{\label{fig:graph-XYZ}A signed weighted graph of $N = 4$ vertices, where $e_{ij}$ is the weight of the edge joining vertices $i$ and $j$ and can be positive or negative. The graph is undirected, so $e_{ji} = e_{ij}$.}
\end{center}
\end{figure}

A network of spin-1/2 particles can be modeled as a graph, where the $N$ vertices are the spins, and the edges indicate which spins can interact. In the Heisenberg model \cite{Heisenberg1928,Baxter1982}, the Hamiltonian consists of two terms, one summing over all pairs of adjacent spins $\{i,j\}$ (i.e., all edges), and another summing over all $N$ spins, given by
\begin{equation}
    \label{eq:Heisenberg}
    \begin{aligned}
        H 
            &= -\frac{1}{2} \sum_{\{i,j\}} \left( J_{ij,x} X_i X_j + J_{ij,y} Y_i Y_j + J_{ij,z} Z_i Z_j \right) \\
            &\quad + \sum_{i=1}^N h_i Z_i,
    \end{aligned}
\end{equation}
where the $J_{ij}$'s are the coupling constants between spins $i$ and $j$ in the $x$, $y$, and $z$ directions, $X_i$, $Y_i$, and $Z_i$ are the Pauli matrices applied to the $i$th spin, and $h_i$ is the strength of the magnetic field at the $i$th spin. For example, for the graph in \fref{fig:graph-XYZ}, the Hamiltonian is
\begin{align*}
    H = &-\frac{1}{2} \big[ 
        \left( J_{12,x} X_1 X_2 + J_{12,y} Y_1 Y_2 + J_{12,z} Z_1 Z_2 \right) \\
        &\quad\quad +\left( J_{23,x} X_2 X_3 + J_{23,y} Y_2 Y_3 + J_{23,z} Z_2 Z_3 \right) \\
        &\quad\quad +\left( J_{24,x} X_2 X_4 + J_{24,y} Y_2 Y_4 + J_{24,z} Z_2 Z_4 \right) \\
        &\quad\quad +\left( J_{34,x} X_3 X_4 + J_{34,y} Y_3 Y_4 + J_{34,z} Z_3 Z_4 \right) \big] \\
        &+ h_1 Z_1 + h_2 Z_2 + h_3 Z_3 + h_4 Z_4.
\end{align*}

A spin-1/2 particle has two eigenstates in $Z$, denoted spin up $\ket{\uparrow}$ and spin down $\ket{\downarrow}$, with corresponding eigenvalues $1$ and $-1$. The Pauli matrices act on these eigenstates of $Z$ by
\begin{align*}
    & X \ket{\uparrow} = \ket{\downarrow}, && Y \ket{\uparrow} = i \ket{\downarrow}, && Z \ket{\uparrow} = \ket{\uparrow}, \\
    & X \ket{\downarrow} = \ket{\uparrow}, && Y \ket{\uparrow} = -i \ket{\downarrow}, && Z \ket{\downarrow} = -\ket{\downarrow}.
\end{align*}

If the spin network contains a spin-up particle at a vertex and spin-down particles at the remaining vertices, then it has a single excitation---the spin-up particle---and will evolve in a subspace---the single-excitation subspace---spanned by basis vectors indicating which vertex the excitation is at. For example, for the graph in \fref{fig:graph-XYZ}, there are four basis vectors, corresponding to whether the excitation is at vertex 1, 2, 3, or 4:
\[
    \ket{1} = \ket{\uparrow\downarrow\downarrow\downarrow},
    \enspace \ket{2} = \ket{\downarrow\uparrow\downarrow\downarrow},
    \enspace \ket{3} = \ket{\downarrow\downarrow\uparrow\downarrow},
    \enspace \ket{4} = \ket{\downarrow\downarrow\downarrow\uparrow}.
\]

It was previously shown that a single excitation in a spin network can effect the Laplacian and adjacency quantum walks \cite{Bose2009}, as well as the signless Laplacian quantum walk \cite{Wong43}, on unweighted graphs. In this paper, we show that the generalized Laplacian quantum walk, which includes the Laplacian, adjacency, and signless Laplacian quantum walks, are effected on weighted graphs as well, and furthermore the weights can be positive or negative. In particular, if the weight of the edge joining vertices $i$ and $j$ is $e_{ij}$ (or, equivalently, $e_{ji}$), then we will now show that when the $x$ and $y$ coupling constants equal the jumping rate times the weight of the edge, i.e., $J_{ij,x} = J_{ij,y} = \gamma e_{ij}$, while the $z$ coupling constant is $1-\alpha$ times this, i.e., $J_{ij,z} = (1-\alpha) \gamma e_{ij}$, and the magnetic fields are all zero, i.e., $h_i = 0$ for all $i$, then the Heisenberg model Hamiltonian \eqref{eq:Heisenberg} in the single-excitation subspace is equal to the generalized Laplacian quantum walk $H_\alpha = -\gamma L_\alpha$. Then, the Laplacian, adjacency, and signless Laplacian quantum walks occur when $\alpha = 0, 1, 2$, respectively.

To begin the proof, with the aforementioned coupling constants and magnetic fields, the Heisenberg model Hamiltonian \eqref{eq:Heisenberg} becomes
\begin{align}
    H 
        &= -\frac{1}{2} \sum_{\{i,j\}} \left( J_{ij,x} X_i X_j + J_{ij,y} Y_i Y_j + J_{ij,z} Z_i Z_j \right) \nonumber \\
        &= -\frac{\gamma}{2} \sum_{\{i,j\}} e_{ij} \left( X_i X_j + Y_i Y_j \right) - (1-\alpha) \frac{\gamma}{2} \sum_{\{i,j\}} e_{ij} Z_i Z_j \nonumber \\
        &= -\frac{\gamma}{2} H_{XY} - (1-\alpha) \frac{\gamma}{2} H_Z, \label{eq:Heisenberg-generalized}
\end{align}
where we have denoted the sums
\[ H_{XY} = \sum_{\{i,j\}} e_{ij} \left( X_i X_j + Y_i Y_j \right) \]
and
\[ H_Z = \sum_{\{i,j\}} e_{ij} Z_i Z_j. \]
Let us focus on each sum separately.

Beginning with $H_{XY}$, as shown in \cite{Wong43}, $X_i X_j + Y_i Y_j$ causes an excitation at vertex $i$ to jump to vertex $j$ with a factor of 2, and vice versa. For example, for the graph in \fref{fig:graph-XYZ},
\begin{align*}
    \left( X_1 X_2 + Y_1 Y_2 \right) \ket{1} &= 2 \ket{2}, \\
    \left( X_1 X_2 + Y_1 Y_2 \right) \ket{2} &= 2 \ket{1}, \\
    \left( X_2 X_3 + Y_2 Y_3 \right) \ket{2} &= 2 \ket{3}, \\
    \left( X_2 X_4 + Y_2 Y_4 \right) \ket{2} &= 2 \ket{4}, \\
    \left( X_2 X_3 + Y_2 Y_3 \right) \ket{3} &= 2 \ket{2}, \\
    \left( X_3 X_4 + Y_3 Y_4 \right) \ket{3} &= 2 \ket{4}, \\
    \left( X_2 X_4 + Y_2 Y_4 \right) \ket{4} &= 2 \ket{2}, \\
    \left( X_3 X_4 + Y_3 Y_4 \right) \ket{4} &= 2 \ket{3}.
\end{align*}
Then,
\begin{align*}
    H_{XY} \ket{1} &= 2 e_{12} \ket{2}, \\
    H_{XY} \ket{2} &= 2 \left( e_{21} \ket{1} + e_{23} \ket{3} + e_{24} \ket{4} \right), \\
    H_{XY} \ket{3} &= 2 \left( e_{32} \ket{2} + e_{34} \ket{4} \right), \\
    H_{XY} \ket{4} &= 2 \left( e_{42} \ket{2} + e_{43} \ket{3} \right).
\end{align*}
Or as a matrix,
\[ H_{XY} = 2 \begin{pmatrix}
    0 & e_{12} & 0 & 0 \\
    e_{21} & 0 & e_{23} & e_{24} \\
    0 & e_{32} & 0 & e_{34} \\
    0 & e_{42} & e_{43} & 0 \\
\end{pmatrix}. \]
The matrix is precisely the adjacency matrix for the signed unweighted graph, and so
\[ H_{XY} = 2A. \]
This is true in general, not just for the example graph in \fref{fig:graph-XYZ}.

Next, for $H_z$, note from \cite{Wong43} that $Z_i Z_j$ flips the sign of $\ket{i}$ and $\ket{j}$ and does nothing to the other single-excitation basis states, i.e., it acts on basis state $\ket{k}$ by
\[ Z_i Z_j \ket{k} = \begin{cases}
    -\ket{k}, & k = i \text{ or } k = j, \\
    \ket{k}, & \text{otherwise}.
\end{cases} \]
Then, for the graph in \fref{fig:graph-XYZ},
\begin{align*}
    H_Z \ket{1}
        &= \left( e_{12} Z_1 Z_2 + e_{23} Z_2 Z_3 \right. \\
        &\quad\enspace \left. + e_{24} Z_2 Z_4 + e_{34} Z_3 Z_4 \right) \ket{1} \\
        &= \left( -e_{12} + e_{23} + e_{24} + e_{34} \right) \ket{1} \\
        &= \left( e_{12} + e_{23} + e_{24} + e_{34} - 2 e_{12}\right) \ket{1} \\
        &= \left( \sum_{\{i,j\}} e_{ij} - 2 e_{12} \right) \ket{1} \\
        &= \left[ |E| - 2 \deg(1) \right] \ket{1},
\end{align*}
where we have denoted $|E|$ as the sum of all the edges of the graph (since for an unweighted graph, each edge has weight 1, and $|E|$ would be the total number of edges) and $\deg(1)$ as the degree of vertex 1. Similarly,
\begin{align*}
    H_Z \ket{2}
        &= \left( \sum_{\{i,j\}} e_{ij} - 2 e_{21} - 2 e_{23} - 2 e_{24} \right) \ket{2} \\
        &= \left[ |E| - 2 \deg(2) \right] \ket{2}, \\
    H_Z \ket{3}
        &= \left( \sum_{\{i,j\}} e_{ij} - 2 e_{32} - 2 e_{34} \right) \ket{3} \\
        &= \left[ |E| - 2 \deg(3) \right] \ket{3}, \\
    H_Z \ket{2}
        &= \left( \sum_{\{i,j\}} e_{ij} - 2 e_{42} - 2 e_{43} \right) \ket{4} \\
        &= \left[ |E| - 2 \deg(4) \right] \ket{4}.
\end{align*}
In general,
\begin{align*}
    H_Z \ket{k}
        &= \left( \sum_{\{i,j\}} e_{ij} - 2 \sum_{\ell \sim k} e_{k\ell} \right) \ket{k} \\
        &= \left( |E| - 2 \deg(k) \right) \ket{k}.
\end{align*}
Then, 
\[ H_Z = |E|I - 2D, \]
where $I$ is the identity matrix and $D$ is the degree matrix. In the Hamiltonian, terms proportional to the identity matrix contribute a global phase, which is physically irrelevant and can be dropped. This can also be interpreted as a rezeroing of energy. Then,
\[ H_Z = -2D. \]

Plugging in $H_{XY} = 2A$ and $H_Z = -2D$ into the Heisenberg model Hamiltonian \eqref{eq:Heisenberg-generalized},
\begin{align*}
    H
        &= -\frac{\gamma}{2} H_{XY} - (1-\alpha) \frac{\gamma}{2} H_Z \\
        &= -\gamma A + (1-\alpha) \gamma D \\
        &= -\gamma (A - D + \alpha D ) \\
        &= -\gamma (L + \alpha D) \\
        &= -\gamma L_\alpha \\
        &= H_\alpha,
\end{align*}
so we have shown that a spin network with a single excitation, with appropriate coupling constants and no external magnetic field, can implement generalized Laplacian quantum walks on signed weighted graphs, which includes Laplacian, adjacency, and signless Laplacian quantum walks when $\alpha = 0, 1, 2$, respectively.


\section{\label{sec:barbell}Searching the Weighted Barbell Graph}

With a physical system in place that motivates the generalized Laplacian quantum walk (i.e., single-excitation spin networks), we now explore an algorithmic consequence of the generalized Laplacian quantum walk by investigating how it searches a weighted barbell graph consisting of two complete graphs of equal size joined by a single signed weighted edge, an example of which was shown in \fref{fig:barbell}. Within each complete graph or clique, the weights of the edges are 1, but the single edge that bridges the two cliques has variable weight $w \in \mathbb{R}$.


\subsection{Algorithm}

A vertex $\ket{a}$ is marked by an oracle, indicated in \fref{fig:barbell} with a double circle. The system $\ket{\psi(t)}$ begins in a uniform superposition over the $N$ vertices,
\[ \ket{\psi(0)} = \frac{1}{\sqrt{N}} \sum_{i=1}^N \ket{i}, \]
and it evolves by Schr\"odinger's equation
\[ i\hbar \frac{d\ket{\psi}}{dt} = H \ket{\psi}, \]
where we will take $\hbar = 1$ throughout this paper, and the Hamiltonian is
\begin{equation}
    \label{eq:H}
    H = -\gamma L_\alpha - \ketbra{a}{a},
\end{equation}
where the first term effects the generalized Laplacian quantum walk with jumping rate $\gamma$, and the second term is a Hamiltonian oracle marking the vertex to be found \cite{CG2004}.

In this paper, we show that the oracle can be implemented in the single-excitation subspace of a spin network through an external magnetic field at the marked vertex of magnitude $1/2$ in the negative $z$ direction, i.e., $h_a = -1/2$ while $h_{i \ne a} = 0$, so the Heisenberg model Hamiltonian is
\[ H = -\gamma L_\alpha - \frac{1}{2} Z_a. \]
This is because $Z_i$ acts on single-excitation basis states by
\[ Z_a \ket{i} = \begin{cases}
    \ket{i}, & i = a, \\
    -\ket{i}, & i \ne a, \\
\end{cases} \]
and so
\[ Z_a = 2\ketbra{a}{a} - I, \]
and since a multiple of the identity matrix can be dropped in the Hamiltonian,
\[ -\frac{1}{2} Z_a = -\ketbra{a}{a}, \]
and so this magnetic field implements the oracle.

Since the search Hamiltonian \eqref{eq:H} is time-independent, the solution to Schr\"odinger's equation is
\[ \ket{\psi(t)} = e^{-iHt} \ket{\psi(0)}. \]
The goal is that when the position of the walker is measured, it is found at the marked vertex, hence solving the search problem, and the probability of this occurring at time $t$ is
\[ p_a(t) = \left| \braket{a}{\psi(t)} \right|^2, \]
which is called the success probability.

In \cite{Wong41}, search on this weighted barbell graph with increasing weights $w \ge 1$ was explored using the Laplacian quantum walk ($\alpha = 0$) and the adjacency quantum walk ($\alpha = 1)$. In both cases, the jumping rate $\gamma$ needed to take a critical value of $2/N$ for the system to appreciably evolve from its initial state. It was shown that with the Laplacian quantum walk, no matter how large the weight $w$, amplitude from one clique could not appreciably move from one clique to the other. It was as if the two cliques were disconnected, with each clique having half the total probability. Since a continuous-time quantum walk searches a complete graph of size $M$ with probability $1$ in time $(\pi/2)\sqrt{M}$, the weighted barbell graph was searched with probability $1/2$ in time $(\pi/2)\sqrt{N/2}$. On the other hand, it was shown that with the adjacency quantum walk, when the weight took a critical value of $w = N/2$, then amplitude was able to move across the bridge, and the success probability reached 82.0\% at time $2.518 \sqrt{N}$. This was improved further by utilizing a two-stage algorithm, where in the first stage, $w = N/2$, and 99.6\% of the probability accumulated in the clique containing the marked vertex (or ``marked clique,'' in contrast to the clique that does not contained the marked vertex, or ``unmarked clique'') at time $3.265 \sqrt{N}$. In the second stage, which ran for an additional time $1.345 \sqrt{N}$), this probability was focused onto the marked vertex by taking some weight that effectively disconnects the cliques (e.g., $w = 1$). Then, the success probability reached 99.6\% in a total time of $4.610 \sqrt{N}$.

In the following subsections, let us explore how the algorithm behaves with real-valued $w$ and for general $\alpha$, beginning with some numerical results followed by analytical proofs.


\subsection{Numerical Simulations}

\begin{figure*}
\begin{center}
    \subfloat{
        \label{fig:prob-time-N1200-neg3}
        \includegraphics{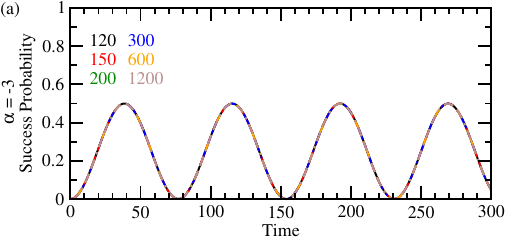}
        \includegraphics{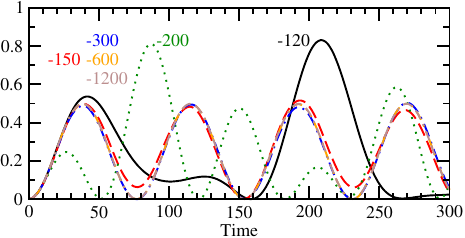}
    }
    
    \subfloat{
        \label{fig:prob-time-N1200-0}
        \includegraphics{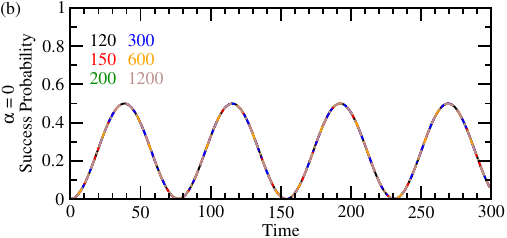}
        \includegraphics{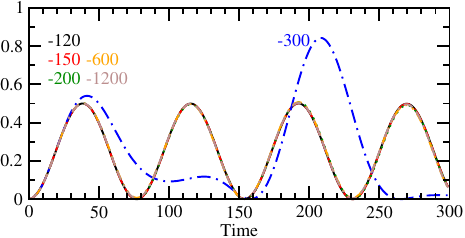}
    }
    
    \subfloat{
        \label{fig:prob-time-N1200-1}
        \includegraphics{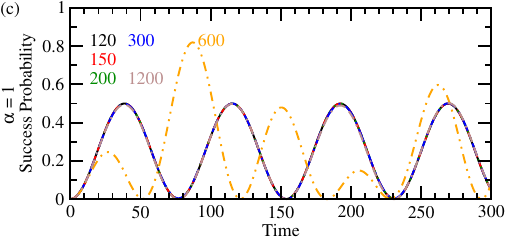}
        \includegraphics{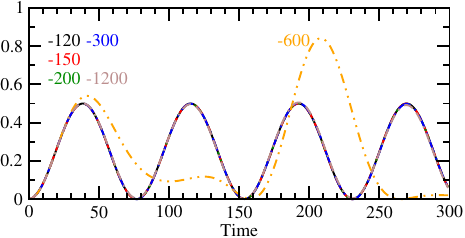}
    }
    
    \subfloat{
        \label{fig:prob-time-N1200-2}
        \includegraphics{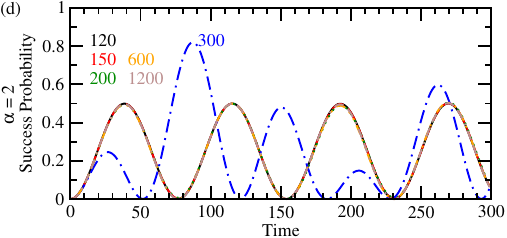}
        \includegraphics{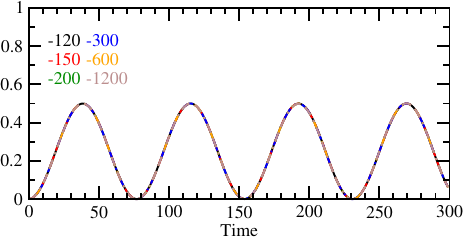}
    }
    
    \subfloat{
        \label{fig:prob-time-N1200-4}
        \includegraphics{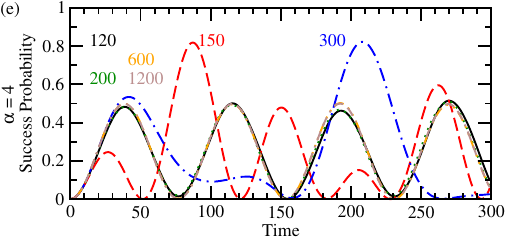}
        \includegraphics{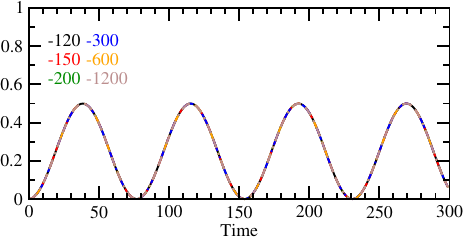}
    }
    \caption{\label{fig:prob-time-N1200}Success probability vs time for search on the weighted barbell graph with $N = 1200$ vertices using the generalized Laplacian quantum walk with (a) $\alpha = -3$, (b) $\alpha = 0$, (c) $\alpha = 1$, (d) $\alpha = 2$, and (e) $\alpha = 4$. In all the left plots, the solid black curve is when the weight of the bridge is $w = 120$, dashed red is $w = 150$, dotted green is $w = 200$, dot-dashed blue is $w = 300$, dot-dot-dashed orange is $w = 600$, and dot-dashed-dashed brown is $w = 1200$. In all the right plots, the respective curves are when the weights are negated, i.e., $w = -120$, $-150$, $-200$, $-300$, $-600$, and $-1200$.}
\end{center}
\end{figure*}

In \fref{fig:prob-time-N1200}, we plot the success probability of the algorithm as it evolves with time for search on the weighted barbell graph of $N = 1200$ vertices. \Fref{fig:prob-time-N1200-neg3} uses the the generalized Laplacian with $\alpha = -3$, and it consists of two plots: a left plot with various positive weights for the bridge, $w = 120$, $150$, $200$, $300$, $600$, and $1200$, and a right plot with the negatives of these weights, i.e., $w = -120$, $-150$, $-200$, $-300$, $-600$, and $-1200$. Most of the curves overlap, reaching a success probability of $1/2$ at time $(\pi/2)\sqrt{N/2} \approx 38.5$ because the graph acts like two disconnected cliques. There are two critical weights with negative values, however, $w = -120$ and $w = -200$, where probability is able to move across the bridge, and a higher success probability is obtained.

In \fref{fig:prob-time-N1200-0}, the success probability is plotted for search with the generalized Laplacian with $\alpha = 0$, i.e., with the standard Laplacian. The left plot shows that positive weights do not help probability move across the bridge, as proved in \cite{Wong41}. The right plot, however, shows that there is a negative critical weight, $w = -300$, where the success probability is improved, and this was not discovered in \cite{Wong41} because that work only considered positive weights.

The generalized Laplacian with $\alpha = 1$, which is the adjacency matrix, was used in \fref{fig:prob-time-N1200-1}. In the left plot, there is a positive critical weight $w = 600$ that boosts the success probability, and it was proved in \cite{Wong41} that this corresponds to a weight of $w = N/2$ and reaches a success probability of $0.820$ (independent of $N$) at a time of $2.518 \sqrt{N} = 87.2$. Ref.~\cite{Wong41} did not consider negative weights, but we do in the right plot, and we see that there is a negative critical weight of $w = -600$ also boosts the success probability.

\Fref{fig:prob-time-N1200-2} is $\alpha = 2$, which is the signless Laplacian. There is only one critical weight that boosts the success probability, $w = 300$.

As a last numerical example, in \fref{fig:prob-time-N1200-4}, the success probability with $\alpha = 4$ is shown, and there are two positive critical weights, $w = 150$ and $w=300$ that boost the success probability.

Note $\alpha$ need not be an integer, and additional simulations show that the $\alpha < 0$ cases all behave similarly: there are two negative critical weights with boosted success probability. The $\alpha = 0$ case (Laplacian) has one negative critical weight. When $0 < \alpha < 2$, there is one positive critical weight and one negative critical weight. When $\alpha = 2$ (signless Laplacian), there is one positive critical weight. When $\alpha > 2$, there are two positive critical weights. Furthermore, for the weighted barbell graph with $N = 1200$ vertices, the boosted success probability either occurs at time $87.2$ or at time $208.2$. Next, we will analytically derive these results, beginning with the critical weights for general $N$ and $\alpha$, followed by the success probabilities and runtimes.


\subsection{Noncritical and Critical Weights}

\begin{figure}
\begin{center}
    \includegraphics{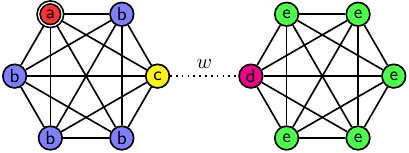}
    \caption{\label{fig:barbell-subspace}The weighted barbell graph of $N = 12$ vertices from \fref{fig:barbell}, with identically evolving vertices are identically colored and labeled.}
\end{center}
\end{figure}

To begin our analysis, as noted in \cite{Wong41}, due to the symmetry of the graph, there are five types of vertices, which are labeled $a$, $b$, $c$, $d$, and $e$ in \fref{fig:barbell-subspace}. Each vertex of the same type evolves identically, so the system evolves in a five-dimensional (5D) subspace spanned by a basis formed by uniform superpositions of each type of vertex:
\begin{gather*}
    \ket{a}, \quad \ket{b} = \frac{1}{\sqrt{N/2-2}} \sum_{i \in b} \ket{i}, \\
    \ket{c}, \quad \ket{d}, \quad \ket{e} = \frac{1}{\sqrt{N/2-1}} \sum_{i \in e} \ket{i}.
\end{gather*}
As given in \cite{Wong41}, in this $\{ \ket{a}, \ket{b}, \ket{c}, \ket{d}, \ket{e} \}$ basis, the initial uniform state is
\begin{equation}
    \label{eq:psi0-5D}
    \ket{\psi(0)} = \frac{1}{\sqrt{N}} \begin{pmatrix} 1 \\ \sqrt{\frac{N}{2} - 2} \\ 1 \\ 1 \\ \sqrt{\frac{N}{2} - 1} \\
    \end{pmatrix},
\end{equation}
the adjacency matrix is
\begin{widetext}
\[ A = \begin{pmatrix}
	0 & \sqrt{\frac{N}{2}-2} & 1 & 0 & 0 \\
	\sqrt{\frac{N}{2}-2} & \frac{N}{2} - 3 & \sqrt{\frac{N}{2}-2} & 0 & 0 \\
	1 & \sqrt{\frac{N}{2}-2} & 0 & w & 0 \\
	0 & 0 & w & 0 & \sqrt{\frac{N}{2}-1} \\
	0 & 0 & 0 & \sqrt{\frac{N}{2}-1} & \frac{N}{2}-2 \\
\end{pmatrix}, \]
and the degree matrix is
\[ D = \begin{pmatrix}
    \frac{N}{2} - 1 & 0 & 0 & 0 & 0 \\
    0 & \frac{N}{2} - 1 & 0 & 0 & 0 \\
    0 & 0 & \frac{N}{2} - 1 + w & 0 & 0 \\
    0 & 0 & 0 & \frac{N}{2} - 1 + w & 0 \\
    0 & 0 & 0 & 0 & \frac{N}{2} - 1 \\
\end{pmatrix}. \]
Now, we note that the degree matrix can be written as
\[ D = \left( \frac{N}{2} - 1 \right) I + \begin{pmatrix}
    0 & 0 & 0 & 0 & 0 \\
    0 & 0 & 0 & 0 & 0 \\
    0 & 0 & w & 0 & 0 \\
    0 & 0 & 0 & w & 0 \\
    0 & 0 & 0 & 0 & 0 \\
\end{pmatrix}, \]
where $I$ is the identity matrix. This will contribute a multiple of the identity matrix to the Hamiltonian, which contributes an indistinguishable, global phase. So, we can drop it, leaving
\begin{equation}
    \label{eq:D}
    D = \begin{pmatrix}
        0 & 0 & 0 & 0 & 0 \\
        0 & 0 & 0 & 0 & 0 \\
        0 & 0 & w & 0 & 0 \\
        0 & 0 & 0 & w & 0 \\
        0 & 0 & 0 & 0 & 0 \\
    \end{pmatrix}.
\end{equation}
Then, the generalized Laplacian $L_\alpha = L + \alpha D = A - D + \alpha D = A + (\alpha-1)D$ is
\[ L_\alpha = \begin{pmatrix}
	0 & \sqrt{\frac{N}{2}-2} & 1 & 0 & 0 \\
	\sqrt{\frac{N}{2}-2} & \frac{N}{2} - 3 & \sqrt{\frac{N}{2}-2} & 0 & 0 \\
	1 & \sqrt{\frac{N}{2}-2} & (\alpha-1)w & w & 0 \\
	0 & 0 & w & (\alpha-1)w & \sqrt{\frac{N}{2}-1} \\
	0 & 0 & 0 & \sqrt{\frac{N}{2}-1} & \frac{N}{2}-2 \\
\end{pmatrix}. \]
Including the oracle $\ketbra{a}{a} = \text{diag} \begin{pmatrix} 1 & 0 & 0 & 0 & 0 \end{pmatrix}$, the search Hamiltonian \eqref{eq:H} is
\begin{equation}
    \label{eq:H-5D}
    H = -\gamma \begin{pmatrix}
	   \frac{1}{\gamma} & \sqrt{\frac{N}{2}-2} & 1 & 0 & 0 \\
	   \sqrt{\frac{N}{2}-2} & \frac{N}{2} - 3 & \sqrt{\frac{N}{2}-2} & 0 & 0 \\
	   1 & \sqrt{\frac{N}{2}-2} & (\alpha-1)w & w & 0 \\
	   0 & 0 & w & (\alpha-1)w & \sqrt{\frac{N}{2}-1} \\
	   0 & 0 & 0 & \sqrt{\frac{N}{2}-1} & \frac{N}{2}-2 \\
    \end{pmatrix}.
\end{equation}
\end{widetext}

We can find the eigenvectors and eigenvalues of this search Hamiltonian \eqref{eq:H-5D} for large $N$ using degenerate perturbation theory, where we first find the eigenvectors and eigenvalues of the leading-order Hamiltonian, and the degeneracies are lifted by the inclusion of next-order corrections. The leading-order Hamiltonian is
\[ 
    H^{(0)} = -\gamma \begin{pmatrix}
	   \frac{1}{\gamma} & 0 & 0 & 0 & 0 \\
	   0 & \frac{N}{2} & 0 & 0 & 0 \\
	   0 & 0 & (\alpha-1)w & w & 0 \\
	   0 & 0 & w & (\alpha-1)w & 0 \\
	   0 & 0 & 0 & 0 & \frac{N}{2} \\
    \end{pmatrix}.
\]
Its eigenvectors and eigenvalues are
\begin{align}
    &\ket{a}, && {-1},\phantom{\frac{1}{2}} \nonumber \\
    &\ket{b}, && \frac{-\gamma N}{2} \nonumber \\
    &\ket{cd_+} = \frac{1}{\sqrt{2}} \left( \ket{c} + \ket{d} \right), && {-\alpha\gamma w}, \label{eq:eigensystem-H0} \\
    &\ket{cd_-} = \frac{1}{\sqrt{2}} \left( \ket{c} - \ket{d} \right), && {-(\alpha-2)\gamma w}, \nonumber \\
    &\ket{e}, && \frac{-\gamma N}{2}. \nonumber
\end{align}
Note that the initial state \eqref{eq:psi0-5D} is approximately $(\ket{b} + \ket{e})/\sqrt{2}$ for large $N$, so for the initial state to evolve to $\ket{a}$, we need eigenvectors of the search Hamiltonian that are linear combinations of $\ket{b}$, $\ket{e}$, and $\ket{a}$. From \eqref{eq:eigensystem-H0}, we see that $\ket{b}$ and $\ket{e}$ are always degenerate eigenvectors of $H^{(0)}$, and they can be made degenerate with $\ket{a}$ by letting $\gamma$ take a critical value of
\begin{equation}
    \label{eq:gamma}
    \gamma_c = \frac{2}{N}.
\end{equation}
For now, we assume the weight of the bridge is such that neither $\ket{cd_+}$ nor $\ket{cd_-}$ are degenerate with $\ket{a}$, $\ket{b}$, and $\ket{e}$. Any linear combination of $\ket{a}$, $\ket{b}$, and $\ket{e}$,
\[ \ket{\psi_{abe}} = \alpha_a \ket{a} + \alpha_b \ket{b} + \alpha_e \ket{e} \]
is an eigenvector of $H^{(0)}$, and to lift their degeneracy, we include the next-order corrections to the search Hamiltonian,
\[
    H^{(1)} = -\gamma \begin{pmatrix}
	   0 & \sqrt{\frac{N}{2}} & 0 & 0 & 0 \\
	   \sqrt{\frac{N}{2}} & 0 & \sqrt{\frac{N}{2}} & 0 & 0 \\
	   0 & \sqrt{\frac{N}{2}} & 0 & 0 & 0 \\
	   0 & 0 & 0 & 0 & \sqrt{\frac{N}{2}} \\
	   0 & 0 & 0 & \sqrt{\frac{N}{2}} & 0 \\
    \end{pmatrix},
\]
and find eigenvectors of $H^{(0)} + H^{(1)}$ of the form $\ket{\psi_{abe}}$. That is, we solve
\[ \left( H^{(0)} + H^{(1)} \right) \ket{\psi_{abe}} = E \ket{\psi_{abe}} \]
with $\gamma = \gamma_c$, which in matrix-vector form is
\[ \begin{pmatrix}
    -1 & -\sqrt{\frac{2}{N}} & 0 & \\
    -\sqrt{\frac{2}{N}} & -1 & 0 \\
    0 & 0 & -1 \\
\end{pmatrix} \begin{pmatrix}
    \alpha_a \\
    \alpha_b \\
    \alpha_e \\
\end{pmatrix} = E \begin{pmatrix}
    \alpha_a \\
    \alpha_b \\
    \alpha_e \\
\end{pmatrix}. \]
Solving this, we get three eigenvectors of $H^{(0)} + H^{(1)}$ when $\gamma = \gamma_c$, and also including $\ket{cd_\pm}$ from \eqref{eq:eigensystem-H0}, the asymptotic eigenvectors and eigenvalues of the search Hamiltonian \eqref{eq:H-5D} are
\begin{align}
    \ket{\psi_0} &= \frac{1}{\sqrt{2}} \left( \ket{a} + \ket{b} \right), && E_0 = -1 - \sqrt{\frac{2}{N}}, \nonumber \\
    \ket{\psi_1} &= \ket{e}, &&E_1 = -1, \nonumber \\
    \ket{\psi_2} &= \frac{1}{\sqrt{2}} \left( \ket{a} - \ket{b} \right), && E_2 = -1 + \sqrt{\frac{2}{N}}, \label{eq:eigensystem-noncritical} \\
    \ket{\psi_3} &= \frac{1}{\sqrt{2}} \left( \ket{c} + \ket{d} \right), && E_3 = -\frac{2\alpha w}{N}, \nonumber \\
    \ket{\psi_4} &= \frac{1}{\sqrt{2}} \left( \ket{c} - \ket{d} \right), && E_4 = -\frac{2(\alpha-2) w}{N}. \nonumber
\end{align}
Depending on the particular values of $\alpha$ and $w$, these eigenvalues may not be sorted. In \sref{subsec:wnoncritical}, we will use these eigenvectors and eigenvalues to prove the evolutions in \fref{fig:prob-time-N1200} where the success probability reaches $1/2$ at time $(\pi/2)\sqrt{N/2} \approx 38.5$. In these cases, probability is unable to cross the bridge from one clique to the other, and so it behaves like search on a complete graph of $N/2$ vertices with half the total probability. These eigenvectors and eigenvalues will also be used to prove the two-stage algorithms in \sref{subsec:wminus}, specifically the second stage of the algorithms in Appendices~\ref{appendix:wminus-twostage-abc} and \ref{appendix:wminus-twostage-ab}.

Returning to the eigenvectors and eigenvalues of $H^{(0)}$ in \eqref{eq:eigensystem-H0}, for amplitude in the unmarked clique to move across to the marked clique, it must go through vertices $\ket{c}$ and $\ket{d}$, so we also want $\ket{a}$, $\ket{b}$, and $\ket{e}$ to be degenerate with $\ket{cd_+}$ and/or $\ket{cd_-}$. It is not possible for $\ket{cd_+}$ and $\ket{cd_-}$ to both be degenerate, so only one can be degenerate at a time. For $\ket{a}$, $\ket{b}$, $\ket{e}$, and $\ket{cd_+}$ to all be degenerate, the weight of the bridge must take a critical value of
\begin{equation}
    \label{eq:wplus}
     w_+ = \frac{1}{\gamma_c \alpha} = \frac{N}{2\alpha}.
\end{equation}
Similarly, for $\ket{a}$, $\ket{b}$, $\ket{e}$, and $\ket{cd_-}$ to all be degenerate, the weight of the bridge must take a critical value of
\begin{equation}
    \label{eq:wminus}
    w_- = \frac{1}{\gamma_c(\alpha-2)} = \frac{N}{2(\alpha-2)}.
\end{equation}
These two critical weights are computed in \tref{table:critical-weights} for a weighted barbell graph of $N = 1200$ vertices for various parameters $\alpha$ of the generalized Laplacian. These are in agreement with \fref{fig:prob-time-N1200}, which showed that $\alpha = -3$ had critical weights of $-200$ and $-120$, $\alpha = 0$ had one critical weight of $-300$, $\alpha = 1$ had two critical weights of $600$ and $-600$, $\alpha = 2$ had one critical weight of $300$, and $\alpha = 4$ had two critical weights of $150$ and $300$. The formulas \eqref{eq:wplus} and \eqref{eq:wminus} also reveal why there are two negative critical weights when $\alpha < 0$, one negative critical weight when $\alpha = 0$, one positive and one negative critical weight when $0 < \alpha < 2$, one positive critical weight when $\alpha = 2$, and two positive critical weights when $\alpha > 2$. It also reveals why we get one evolution when $w = N/(2\alpha)$ and another when $w = N/[2(\alpha-2)]$, i.e., either $\ket{cd_+}$ or $\ket{cd_-}$ is degenerate with $\ket{a}$, $\ket{b}$, and $\ket{e}$, respectively. The other weights in the plots are noncritical weights, where probability cannot appreciably cross the bridge.

\begin{table}
\caption{\label{table:critical-weights}Critical weights for search on a weighted barbell graph of $N = 1200$ vertices with a generalized quantum walk $L_\alpha$, which permits amplitude to flow between the cliques.}
\begin{ruledtabular}
\begin{tabular}{ccc}
    $\alpha$ & $\displaystyle w_+ = \frac{N}{2\alpha}$ & $\displaystyle w_- = \frac{N}{2(\alpha-2)}$ \\
    \hline
    -5 & -120 & -600/7 \\
    -4 & -150 & -100 \\
    -3 & -200 & -120 \\
    -2 & -300 & -150 \\
    -1 & -600 & -200 \\
    0 & \text{undefined} & -300 \\
    1 & 600 & -600 \\
    2 & 300 & \text{undefined} \\
    3 & 200 & 600 \\
    4 & 150 & 300 \\
    5 & 120 & 200 \\
\end{tabular}
\end{ruledtabular}
\end{table}

So, when $\gamma = \gamma_c$ and $w = w_\pm$, any linear combination of $\ket{a}$, $\ket{b}$, $\ket{e}$, and $\ket{cd_\pm}$,
\[ \ket{\psi_\pm} = \alpha_a \ket{a} + \alpha_b \ket{b} + \alpha_{cd} \ket{cd_\pm} + \alpha_e \ket{e}, \]
is an eigenvector of $H^{(0)}$. To lift the degeneracy, we include the next-order corrections to the search Hamiltonian and find eigenvectors of $H^{(0)} + H^{(1)}$ of the form $\ket{\psi_\pm}$. That is, we solve
\[ \left( H^{(0)} + H^{(1)} \right) \ket{\psi_\pm} = E \ket{\psi_\pm} \]
with $\gamma = \gamma_c$ and $w = w_\pm$, which in matrix-vector form is
\[ \begin{pmatrix}
    -1 & -\sqrt{\frac{2}{N}} & 0 & 0 \\
    -\sqrt{\frac{2}{N}} & -1 & -\frac{1}{\sqrt{N}} & 0 \\
    0 & -\frac{1}{\sqrt{N}} & -1 & \mp\frac{1}{\sqrt{N}} \\
    0 & 0 & \mp\frac{1}{\sqrt{N}} & -1 \\
\end{pmatrix} \begin{pmatrix}
    \alpha_a \\
    \alpha_b \\
    \alpha_{cd} \\
    \alpha_{e} \\
\end{pmatrix} = E \begin{pmatrix}
    \alpha_a \\
    \alpha_b \\
    \alpha_{cd} \\
    \alpha_{e} \\
\end{pmatrix}. \]
Solving this, the (unnormalized) eigenvectors of $H^{(0)} + H^{(1)}$ when $\gamma = \gamma_c$ and $w = w_\pm$ include
\begin{equation}
    \label{eq:eigenvectors}
    \begin{aligned}
        \psi_{\pm,0}
            &= \sqrt{2 + \sqrt{2}} \ket{a} + \left( 1 + \sqrt{2} \right) \ket{b} \\
            &\quad + \sqrt{2 + \sqrt{2}} \ket{cd_\pm} \pm \ket{e}, \\
        \psi_{\pm,1} 
            &= -\sqrt{2 - \sqrt{2}} \ket{a} + \left( 1 - \sqrt{2} \right) \ket{b} \\
            &\quad + \sqrt{2 - \sqrt{2}} \ket{cd_\pm} \pm \ket{e}, \\
        \psi_{\pm,2} 
            &= \sqrt{2 - \sqrt{2}} \ket{a} + \left( 1 - \sqrt{2} \right) \ket{b} \\
            &\quad - \sqrt{2 - \sqrt{2}} \ket{cd_\pm} \pm \ket{e}, \\
        \psi_{\pm,3} 
            &= -\sqrt{2 + \sqrt{2}} \ket{a} + \left( 1 + \sqrt{2} \right) \ket{b} \\
            &\quad - \sqrt{2 + \sqrt{2}} \ket{cd_\pm} \pm \ket{e},
    \end{aligned}
\end{equation}
and their corresponding eigenvalues are
\begin{equation}
    \label{eq:eigenvalues}
    \begin{aligned}
        E_{\pm,0} &= -1 - \sqrt{\frac{2 + \sqrt{2}}{N}}, \\
        E_{\pm,1} &= -1 - \sqrt{\frac{2 - \sqrt{2}}{N}}, \\
        E_{\pm,2} &= -1 + \sqrt{\frac{2 - \sqrt{2}}{N}}, \\
        E_{\pm,3} &= -1 + \sqrt{\frac{2 + \sqrt{2}}{N}}.
    \end{aligned}
\end{equation}
Using these eigenvectors and eigenvalues, we will determine the evolution of the algorithm when $w = w_+$ in \sref{subsec:wplus} and when $w = w_-$ in \sref{subsec:wminus}. But first, we next prove the behavior when $w$ takes neither of these critical values in \sref{subsec:wnoncritical}.


\subsection{\label{subsec:wnoncritical}Behavior when \texorpdfstring{$w$}{w} is noncritical}

When $w$ is noncritical and $N$ is large, the relevant eigenvectors and eigenvalues of the search Hamiltonian are given by \eqref{eq:eigensystem-noncritical}. Expressing the initial state in terms of these eigenvectors, the system evolves to
\begin{align*}
    \ket{\psi(t)} 
        &= e^{-iHt} \ket{\psi(0)} \\
        &\approx e^{-iHt} \frac{1}{\sqrt{2}} \left( \ket{b} + \ket{e} \right) \\
        &= e^{-iHt} \frac{1}{\sqrt{2}} \left[ \frac{1}{\sqrt{2}} \left( \ket{\psi_0} - \ket{\psi_2} \right) + \ket{\psi_1} \right] \\
        &= \frac{1}{2} \left( e^{-iE_0t} \ket{\psi_0} - e^{-iE_2t} \ket{\psi_2} \right) + \frac{1}{\sqrt{2}} e^{-iE_1t} \ket{\psi_1} \\
        &= \frac{1}{2} \left[ e^{-i(-1-\sqrt{2/N})t} \frac{1}{\sqrt{2}} \left( \ket{a} + \ket{b} \right) \right. \\
        &\quad\quad\enspace \left. - e^{-i(-1+\sqrt{2/N})t} \frac{1}{\sqrt{2}} \left( \ket{a} - \ket{b} \right) \right] + \frac{1}{\sqrt{2}} e^{it} \ket{e} \\
        &= \frac{e^{it}}{2\sqrt{2}} \left[ \left( e^{i\sqrt{2/N}t} - e^{-i\sqrt{2/N}t} \right) \ket{a} \right. \\
        &\quad\quad\quad\enspace \left. + \left( e^{i\sqrt{2/N}t} + e^{-i\sqrt{2/N}t} \right) \ket{b} \right] + \frac{1}{\sqrt{2}} e^{it} \ket{e} \\
        &= \frac{e^{it}}{\sqrt{2}} \left[ i\sin \left( \sqrt{\frac{2}{N}} t \right) \ket{a} + \cos \left( \sqrt{\frac{2}{N}} t \right) \ket{b} \right] \\
        &\quad + \frac{1}{\sqrt{2}} e^{it} \ket{e}.
\end{align*}
Taking the norm-square of each amplitude, the probability at each type of vertex is
\begin{align*}
    p_a(t) &= \frac{1}{2} \sin^2 \left( \sqrt{\frac{2}{N}} t \right), \\
    p_b(t) &= \frac{1}{2} \cos^2 \left( \sqrt{\frac{2}{N}} t \right), \\
    p_c(t) &= 0, \\
    p_d(t) &= 0, \\
    p_e(t) &= \frac{1}{2}.
\end{align*}
Then, the success probability $p_a(t)$ reaches a maximum value of $1/2$ when $\sqrt{2/N}t = \pi/2$, i.e., when $t = (\pi/2)\sqrt{N/2}$, and this agrees with \fref{fig:prob-time-N1200}, where for noncritical weights, the success probability reaches $1/2$ at time $(\pi/2)\sqrt{1200/2} \approx 38.5$. This is summarized in the first column of \tref{table:summary}. This proves that when the weight is noncritical, probability is unable to cross the bridge, and the probability in the marked clique moves between the $a$ and $b$ vertices.

\begin{table*}
\caption{\label{table:summary}For large $N$, summary of search on the weighted barbell graph of $N$ vertices with $\gamma = \gamma_c = 2/N$ and noncritical and critical weight bridges. N/A stands for not applicable, as noncritical weights and the critical weight $w = w_-$ do not have two-stage algorithms.}
\begin{ruledtabular}
\begin{tabular}{c|ccc}
    Weight & Noncritical & $\displaystyle w = w_+ = \frac{N}{2\alpha}$ & $\displaystyle w = w_- = \frac{N}{2(\alpha-2)}$ \\
    \hline
    Single-Stage Runtime & $\displaystyle \frac{\pi}{2\sqrt{2}} \sqrt{N} \approx 1.111 \sqrt{N}$ & $2.518 \sqrt{N}$ (transcendental) & $\displaystyle \frac{5\pi\sqrt{N}}{\sqrt{2+\sqrt{2}} + \sqrt{2-\sqrt{2}}} \approx 6.011 \sqrt{N}$ \\
    Single-Stage Success Probability & $1/2 = 0.5$ & $0.820$ (transcendental) & $\displaystyle \frac{2+\sqrt{2}}{4} \sin^2 \left( \frac{5\pi}{\sqrt{2}} \right) \approx 0.843$ \\
    \hline
    Two-Stage Runtime & N/A & $4.610 \sqrt{N}$ & N/A \\
    Two-Stage Success Probability & N/A & 0.996 & N/A \\
\end{tabular}
\end{ruledtabular}
\end{table*}


\subsection{\label{subsec:wplus}Behavior when \texorpdfstring{$w = w_+$}{w = w+}}

When $w = w_+ = N/(2\alpha)$, the eigenvectors [the $\psi_+$'s in \eqref{eq:eigenvectors}] and eigenvalues \eqref{eq:eigenvalues} are identical to those reported in \cite{Wong41} for search with the adjacency quantum walk ($\alpha = 1$), and since these eigenvectors and eigenvalues do not depend on $\alpha$, the asymptotic behavior of the search algorithm is the same for all $\alpha$ with $w = w_+ = N/(2\alpha)$ (except $\alpha = 0$, of course, for which this critical weight does not exist). So, all of the asymptotic results from the adjacency matrix in \cite{Wong41} carry over, including the one-stage algorithm where the success probability reaches 82.0\% (independent of $N$) at time $2.518 \sqrt{N}$, where the numbers are solutions to transcendental equations. This is confirmed in \fref{fig:prob-time-N1200}, where the success probability roughly reaches 82.0\% at time $2.518 \sqrt{1200} \approx 87.2$ when $w = w_+$. This is summarized in the top half of the second column of \tref{table:summary}.

\begin{figure}
\begin{center}
    \subfloat[] {
        \includegraphics{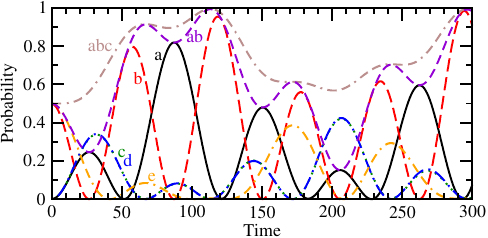}
        \label{fig:prob-time-N1200-4-w150-types}
    }
    
    \subfloat[] {
        \includegraphics{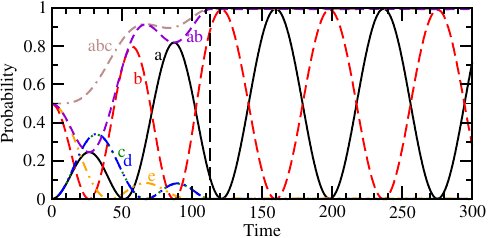}
        \label{fig:prob-time-N1200-4-w150-twostage}
    }
    \caption{Search on the weighted barbell graph with $N = 1200$ vertices using the generalized Laplacian quantum walk with $\alpha = 4$. The solid black curve is the probability at the marked $a$ vertex, i.e., the success probability. The dashed red curve is the probability at the $b$ vertices, dotted green is $c$, dot-dashed blue is $d$ (and overlaps with the dotted green $c$ curve), and dot-dot-dashed orange is $e$. The dot-dashed-dashed brown curve is the probability in the marked clique, i.e., the probability in vertices $a$, $b$, or $c$, and the short dashed violet curve is the probability in vertices $a$ and $b$. In (a), $w = w_+ = N/(2\alpha) = 150$ is used throughout. In (b), $w = w_+ = N/(2\alpha) = 150$ is used left of the vertical dashed line, and $w = 1$ is used right of the vertical dashed line.}
\end{center}
\end{figure}

Also carrying over from the adjacency quantum walk in \cite{Wong41} is a two-stage algorithm, which can be motivated from \fref{fig:prob-time-N1200-4-w150-types}, which plots the probability in each type of vertex $a$, $b$, $c$, $d$, and $e$, as well as the probability in the marked clique (i.e., in vertices $a$, $b$, and $c$) and the probability at vertices $a$ and $b$, when searching a weighted barbell graph of $N = 1200$ vertices using the generalized Laplacian with $\alpha = 4$. Here, $w = w_+$ is used throughout, and we see that around $t = 113.1$, nearly all of the probability is in the marked clique. This accumulation of probability in the marked clique is the first stage of the algorithm, and it is followed by a second stage with a noncritical weight (such as $w = 1$) that focuses the probability onto the marked vertex. Analytically \cite{Wong41}, the critical weight $w = w_+$ is used for time $3.265 \sqrt{N}$, and then a noncritical weight is used for an additional time $1.345 \sqrt{N}$, resulting in the success probability reaching 99.6\% (independent of $N$) in a total time of $4.610 \sqrt{N}$, where the numbers are solutions to transcendental equations. An example of the two-stage algorithm is shown in \fref{fig:prob-time-N1200-4-w150-twostage}. The vertical dashed line at time $3.265 \sqrt{1200} \approx 113.1$ separates the two stages of the algorithm. For the first stage of the algorithm, $w = w_+ = 1200/(2\cdot4) = 150$, and the probability in the marked clique roughly reaches 99.6\%. In the second stage of the algorithm, the noncritical weight $w = 1$ is used, and the success probability roughly reaches 99.6\% at time $4.610 \sqrt{1200} \approx 159.7$. The two-stage algorithm  is summarized in the bottom half of the second column of \tref{table:summary}.

Although the two-stage algorithm in \cite{Wong41} was derived by maximizing the probability in the marked clique in the first stage, in this paper, we note that what really matters is maximizing the probability in the $a$ and $b$ vertices, without regard for the $c$ vertex. This is because in the second stage, with a noncritical weight, the probability only moves between the $a$ and $b$ vertices, while the probability at the $c$ vertex is stuck. So, maximizing the probability in the marked clique is detrimental if it causes a significant amount of probability to build up, and then become stuck, at the $c$ vertex. Fortunately, as proved in Appendix~\ref{appendix:wplus-twostage-ab}, when $w = w_+$, the time at which the probability at the marked clique is maximized is the same as the time at which the probability in the $a$ and $b$ vertices is maximized. This is also seen in \fref{fig:prob-time-N1200-4-w150-types} and \fref{fig:prob-time-N1200-4-w150-twostage}, since the dot-dashed-dashed brown curve corresponding to the probability in the marked clique peaks at the same time as the short dashed violet curve corresponding to the probability in vertices $a$ and $b$. This will not be true for $w = w_-$; as we will see in \sref{subsec:wminus}, these probabilities will peak at different times, and a two-stage algorithm that maximizes the probability in the marked clique will have a lower success probability than one that maximizes the probability in $a$ and $b$.


\subsection{\label{subsec:wminus}Behavior when \texorpdfstring{$w = w_-$}{w = w-}}

When $w = w_- = N/[2(\alpha-2)]$, the eigenvectors of $H^{(0)} + H^{(1)}$ are the $\psi_-$'s in \eqref{eq:eigenvectors}. Then, we can express the initial state \eqref{eq:psi0-5D} as a linear combination of them:
\begin{align*}
   \ket{\psi(0)}
        &\approx \frac{1}{\sqrt{2}} \left( \ket{b} + \ket{e} \right) \displaybreak[0] \\
        &= \frac{2-\sqrt{2}}{8} \left( \psi_{-,0} + \psi_{-,3} \right) - \frac{2+\sqrt{2}}{8} \left( \psi_{-,1} + \psi_{-,2} \right). 
\end{align*}
Then the state at time $t$ is
\begin{widetext}
\begin{align}
    \ket{\psi(t)} 
        &= e^{-iHt} \ket{\psi(0)} \nonumber \displaybreak[0] \\
        &\approx e^{-iHt} \left[ \frac{2-\sqrt{2}}{8} \left( \psi_{-,0} + \psi_{-,3} \right) - \frac{2+\sqrt{2}}{8} \left( \psi_{-,1} + \psi_{-,2} \right) \right] \nonumber \displaybreak[0] \\
        &= \frac{2-\sqrt{2}}{8} \left( e^{-iE_0t} \psi_{-,0} + e^{-iE_3t} \psi_{-,3} \right) - \frac{2+\sqrt{2}}{8} \left( e^{-iE_1t} \psi_{-,1} +e^{-iE_2t} \psi_{-,2} \right) \nonumber \displaybreak[0] \\
        &= \frac{2-\sqrt{2}}{8} \left\{ e^{-i\left(-1-\sqrt{\frac{2+\sqrt{2}}{N}}\right)t} \left[ \sqrt{2+\sqrt{2}} \ket{a} + \left( 1+\sqrt{2} \right) \ket{b} + \sqrt{2+\sqrt{2}} \ket{cd_-} - \ket{e} \right] \right. \nonumber \\
            &\quad\quad\quad\quad\quad + \left. e^{-i\left(-1+\sqrt{\frac{2+\sqrt{2}}{N}}\right)t} \left[ -\sqrt{2+\sqrt{2}} \ket{a} + \left( 1+\sqrt{2} \right) \ket{b} - \sqrt{2+\sqrt{2}}\ket{cd_-} -\ket{e} \right] \right\} \nonumber \displaybreak[0] \\
            &\quad- \frac{2+\sqrt{2}}{8} \left\{ e^{-i\left(-1-\sqrt{\frac{2-\sqrt{2}}{N}}\right)t} \left[ -\sqrt{2-\sqrt{2}} \ket{a} + \left( 1-\sqrt{2} \right) \ket{b} + \sqrt{2-\sqrt{2}} \ket{cd_-} -\ket{e} \right] \right. \nonumber \\
            &\quad\quad\quad\quad\quad\quad + \left. e^{-i\left(-1+\sqrt{\frac{2-\sqrt{2}}{N}}\right)t} \left[ \sqrt{2-\sqrt{2}} \ket{a} + \left( 1-\sqrt{2} \right) \ket{b} - \sqrt{2-\sqrt{2}} \ket{cd_-} - \ket{e} \right] \right\} \nonumber \displaybreak[0] \\
        &= e^{it} \left[ \frac{2-\sqrt{2}}{8} \left( e^{i\sqrt{\frac{2+\sqrt{2}}{N}}t} - e^{-i\sqrt{\frac{2+\sqrt{2}}{N}}t} \right) \sqrt{2+\sqrt{2}} - \frac{2+\sqrt{2}}{8} \left( -e^{i\sqrt{\frac{2-\sqrt{2}}{N}}t} + e^{-i\sqrt{\frac{2-\sqrt{2}}{N}}t} \right) \sqrt{2-\sqrt{2}} \right] \ket{a} \nonumber \\
            &\quad+ e^{it} \left[ \frac{2-\sqrt{2}}{8} \left( e^{i\sqrt{\frac{2+\sqrt{2}}{N}}t} + e^{-i\sqrt{\frac{2+\sqrt{2}}{N}}t} \right) \left( 1+\sqrt{2} \right) - \frac{2+\sqrt{2}}{8} \left( e^{i\sqrt{\frac{2-\sqrt{2}}{N}}t} + e^{i\sqrt{\frac{2-\sqrt{2}}{N}}t} \right) \left( 1-\sqrt{2} \right) \right] \ket{b} \nonumber \\
            &\quad+ e^{it} \left[ \frac{2-\sqrt{2}}{8} \left( e^{i\sqrt{\frac{2+\sqrt{2}}{N}}t} - e^{-i\sqrt{\frac{2+\sqrt{2}}{N}}t} \right) \sqrt{2+\sqrt{2}} - \frac{2+\sqrt{2}}{8} \left( e^{i\sqrt{\frac{2-\sqrt{2}}{N}}t} - e^{-i\sqrt{\frac{2-\sqrt{2}}{N}}t} \right) \sqrt{2-\sqrt{2}} \right] \ket{cd_-} \nonumber \\
            &\quad+ e^{it} \left[ \frac{2-\sqrt{2}}{8} \left( -e^{i\sqrt{\frac{2+\sqrt{2}}{N}}t} - e^{-i\sqrt{\frac{2+\sqrt{2}}{N}}t} \right) - \frac{2+\sqrt{2}}{8} \left( -e^{i\sqrt{\frac{2-\sqrt{2}}{N}}t} - e^{-i\sqrt{\frac{2-\sqrt{2}}{N}}t} \right) \right] \ket{e} \nonumber \displaybreak[0] \\
        &= \frac{ie^{it}}{2\sqrt{2}} \left[ \sqrt{2-\sqrt{2}} \sin \left( \sqrt{\frac{2+\sqrt{2}}{N}}t \right) + \sqrt{2+\sqrt{2}} \sin \left( \sqrt{\frac{2-\sqrt{2}}{N}}t \right) \right] \ket{a} \nonumber \\
            &\quad+ \frac{e^{it}}{2\sqrt{2}} \left[ \cos \left (\sqrt{\frac{2+\sqrt{2}}{N}}t \right) + \cos \left( \sqrt{\frac{2-\sqrt{2}}{N}}t \right) \right] \ket{b} \nonumber \\
            &\quad+ \frac{ie^{it}}{2\sqrt{2}} \left[ \sqrt{2-\sqrt{2}} \sin \left( \sqrt{\frac{2+\sqrt{2}}{N}}t \right) - \sqrt{2+\sqrt{2}} \sin \left( \sqrt{\frac{2-\sqrt{2}}{N}}t \right) \right] \ket{cd_-} \nonumber \\
            &\quad+\frac{e^{it}}{4} \left[ \left( -2+\sqrt{2} \right) \cos \left( \sqrt{\frac{2+\sqrt{2}}{N}}t \right) + \left( 2+\sqrt{2} \right) \cos \left( \sqrt{\frac{2-\sqrt{2}}{N}}t \right) \right] \ket{e}. \label{eq:state-at-time-t}
\end{align}
Using $\ket{cd_-} = \left( \ket{c} - \ket{d} \right) / \sqrt{2}$ and taking the norm-square of each amplitude, the probability of measuring the particle to be at each type of vertex is
\begin{align}
    p_a(t) &= \frac{1}{8} \left[ \sqrt{2-\sqrt{2}} \sin \left( \sqrt{\frac{2+\sqrt{2}}{N}}t \right) + \sqrt{2+\sqrt{2}} \sin \left( \sqrt{\frac{2-\sqrt{2}}{N}}t \right) \right]^2
    , \nonumber \displaybreak[0] \\
    p_b(t) &= \frac{1}{8} \left[ \cos \left (\sqrt{\frac{2+\sqrt{2}}{N}}t \right) + \cos \left( \sqrt{\frac{2-\sqrt{2}}{N}}t \right) \right]^2, \nonumber \displaybreak[0] \\
    p_{c}(t) &= \frac{1}{16} \left[ \sqrt{2-\sqrt{2}} \sin \left( \sqrt{\frac{2+\sqrt{2}}{N}}t \right) - \sqrt{2+\sqrt{2}} \sin \left( \sqrt{\frac{2-\sqrt{2}}{N}}t \right) \right]^2, \label{eq:probs} \displaybreak[0] \\
    p_{d}(t) &= \frac{1}{16} \left[ \sqrt{2-\sqrt{2}} \sin \left( \sqrt{\frac{2+\sqrt{2}}{N}}t \right) - \sqrt{2+\sqrt{2}} \sin \left( \sqrt{\frac{2-\sqrt{2}}{N}}t \right) \right]^2, \nonumber \displaybreak[0] \\
    p_e(t) &= \frac{1}{16} \left[ \left( -2+\sqrt{2} \right) \cos \left( \sqrt{\frac{2+\sqrt{2}}{N}}t \right) + \left( 2+\sqrt{2} \right) \cos \left( \sqrt{\frac{2-\sqrt{2}}{N}}t \right) \right]^2. \nonumber
\end{align}
Note that the success probability $p_a(t)$ has strong agreement with \fref{fig:prob-time-N1200} when $N = 1200$ and $w = w_-$. To find the time at which the success probability peaks, we set the derivative of $p_a(t)$ equal to $0$ and solve for $t$. The derivative of the success probability is
\begin{align*}
    \frac{dp_a}{dt}
        &= \frac{1}{4} \left[ \sqrt{2-\sqrt{2}} \sin \left( \sqrt{\frac{2+\sqrt{2}}{N}}t \right) + \sqrt{2+\sqrt{2}} \sin \left( \sqrt{\frac{2-\sqrt{2}}{N}}t \right) \right] \\
        &\quad \times \left[ \sqrt{2-\sqrt{2}} \sqrt{\frac{2+\sqrt{2}}{N}} \cos \left( \sqrt{\frac{2+\sqrt{2}}{N}}t \right) + \sqrt{2+\sqrt{2}} \sqrt{\frac{2-\sqrt{2}}{N}} \cos \left( \sqrt{\frac{2-\sqrt{2}}{N}}t \right) \right] \displaybreak[0] \\
        &= \frac{1}{2\sqrt{2}\sqrt{N}} \left[ \sqrt{2-\sqrt{2}} \sin \left( \sqrt{\frac{2+\sqrt{2}}{N}}t \right) + \sqrt{2+\sqrt{2}} \sin \left( \sqrt{\frac{2-\sqrt{2}}{N}}t \right) \right] \\
        &\quad \times \left[ \cos \left( \sqrt{\frac{2+\sqrt{2}}{N}}t \right) + \cos \left( \sqrt{\frac{2-\sqrt{2}}{N}}t \right) \right]. \label{eq:dpa}
\end{align*}
\end{widetext}
Setting this equal to zero, the relevant maximum comes from the cosine terms, i.e.,
\[ \cos \left( \sqrt{\frac{2+\sqrt{2}}{N}}t \right) + \cos \left( \sqrt{\frac{2-\sqrt{2}}{N}}t \right) = 0, \]
which becomes
\[ \cos \left( \sqrt{\frac{2+\sqrt{2}}{N}}t \right) = - \cos \left( \sqrt{\frac{2-\sqrt{2}}{N}}t \right), \]
or
\[ \cos \left( \sqrt{\frac{2+\sqrt{2}}{N}}t \right) = \cos \left( k\pi - \sqrt{\frac{2-\sqrt{2}}{N}}t \right), \]
for odd integer $k = 1, 3, 5, \dots$. Taking the inverse cosine of each side,
\[ \sqrt{\frac{2+\sqrt{2}}{N}}t = k\pi - \sqrt{\frac{2-\sqrt{2}}{N}}t, \]
and solving for $t$, we get
\[ t = \frac{k \pi \sqrt{N}}{\sqrt{2+\sqrt{2}} + \sqrt{2-\sqrt{2}}}. \]
The relevant maximum occurs when $k = 5$, so the runtime is
\begin{equation}
    \label{eq:runtime}
    t_* = \frac{5\pi\sqrt{N}}{\sqrt{2+\sqrt{2}} + \sqrt{2-\sqrt{2}}} \approx 6.011 \sqrt{N}.
\end{equation}
For example, when $N = 1200$, the runtime is $t_* \approx 208.2$, in strong agreement with \fref{fig:prob-time-N1200} when $w = w_-$. Note $k = 1,3$ respectively yield the local maxima at $t = 41.6, 124.9$ in \fref{fig:prob-time-N1200} when $w = w_-$, which is why $k = 5$ was chosen in \eqref{eq:runtime}. Plugging $t_*$ into $p_a(t)$ and letting $A = \sqrt{2+\sqrt{2}}$ and $B = \sqrt{2 - \sqrt{2}}$, the success probability reaches a height of
\[ p_a(t_*) = \frac{1}{8} \left[ B \sin \left( \frac{5\pi A}{A+B} \right) + A \sin \left( \frac{5\pi B}{A+B} \right) \right]^2. \]
Since $5\pi B/(A+B) = 5\pi - 5\pi A/(A+B)$, the two sines are equal, and so
\[ p_a(t_*) = \frac{(A+B)^2}{8} \sin^2 \left( \frac{5\pi A}{A+B} \right). \]
Since $(A+B)^2 = 2(2+\sqrt{2})$ and $A/(A+B) = 1/\sqrt{2}$, this simplifies to
\begin{equation}
    \label{eq:prob}
    p_a(t_*) = \frac{2+\sqrt{2}}{4} \sin^2 \left( \frac{5\pi}{\sqrt{2}} \right) \approx 0.843,
\end{equation}
which is in strong agreement with \fref{fig:prob-time-N1200} when $w = w_-$.

Note that in \sref{subsec:wplus}, when $w = w_+$, the runtime and success probability were solutions to transcendental equations \cite{Wong41}, but here when $w = w_-$, we got exact analytical expressions \eqref{eq:runtime} and \eqref{eq:prob}.

\begin{figure}
\begin{center}
    \subfloat[] {
        \includegraphics{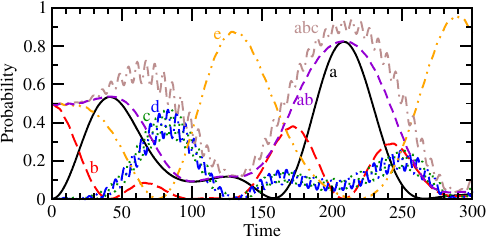}
        \label{fig:prob-time-N1200-4-w300-types}
    }
    
    \subfloat[] {
        \includegraphics{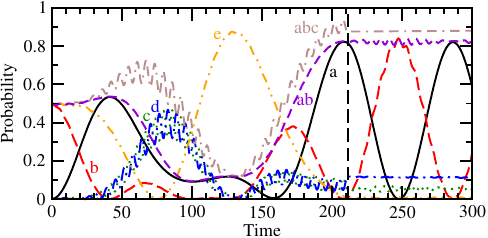}
        \label{fig:prob-time-N1200-4-w300-twostage-abc}
    }
    
    \subfloat[] {
        \includegraphics{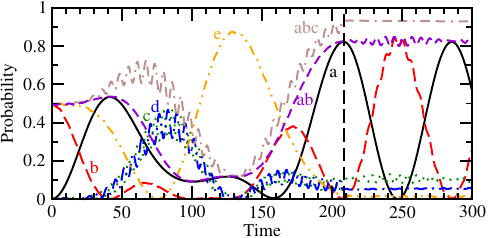}
        \label{fig:prob-time-N1200-4-w300-twostage-ab}
    }
    \caption{Search on the weighted barbell graph with $N = 1200$ vertices using the generalized Laplacian quantum walk with $\alpha = 4$. The solid black curve is the probability at the marked $a$ vertex, i.e., the success probability. The dashed red curve is the probability at the $b$ vertices, dotted green is $c$, dot-dashed blue is $d$, and dot-dot-dashed orange is $e$. The dot-dashed-dashed brown curve is the probability in the marked clique, i.e., the probability in vertices $a$, $b$, or $c$, and the short dashed violet curve is the probability in vertices $a$ and $b$. In (a), $w = w_- = N/[2(\alpha-2)] = 300$ is used throughout. In (b) and (c), $w = w_+ = N/[2(\alpha-2)] = 300$ is used left of the vertical dashed line, and $w = 1$ is used right of the vertical dashed line. In (b), the vertical dashed line was chosen to maximizes the probability in the marked clique, and in (c), it maximizes the probability in vertices $a$ and $b$.}
\end{center}
\end{figure}

Now, we explore potential two-stage algorithms, where $w = w_-$ is used for the first stage, and a noncritical weight, say $w = 1$, is used for the second stage. To motivate the transition time, \fref{fig:prob-time-N1200-4-w300-types} plots the probability in each type of vertex, the marked clique, and the $a$ and $b$ vertices. Whereas in \sref{subsec:wplus}, the probability in the marked clique and the probability in the $a$ and $b$ vertices reached their maxima at the same time, here, the probability in the marked clique peaks a little later, around $t = 211.2$, than the probability in the $a$ and $b$ vertices, which peaks around $t = 208.2$. So, there are two different times that we could transition from the first stage to the second stage of the algorithm.

If we take the transition time to maximize the probability in the marked clique, i.e., around $t = 211.1$, then we get \fref{fig:prob-time-N1200-4-w300-twostage-abc} for the evolution of the two-stage algorithm. For large $N$, we analytically prove in Appendix~\ref{appendix:wminus-twostage-abc} that the success probability reaches $0.840$, which is worse than the single-stage algorithm's $0.843$ \eqref{eq:prob}, and it takes more time. So, one would just use the single-stage algorithm over this two-stage.

If we instead take the transition time to maximize the probability in the $a$ and $b$ vertices, then we get \fref{fig:prob-time-N1200-4-w300-twostage-ab} for the two-stage algorithm. For large $N$, as we analytically prove in Appendix~\ref{appendix:wminus-twostage-ab}, the transition time is the same  as the runtime of the single-stage algorithm \eqref{eq:runtime}, and the success probability of the two-stage algorithm is the same as the single-stage algorithm's $0.843$ \eqref{eq:prob}. Then, it does not make sense to run the algorithm for a second stage, which takes more time, if all it does is reach the same success probability.

Comparing these two-stage algorithms, even though the success probabilities of $0.840$ and $0.843$ only differ by $0.003$, it shows that maximizing the probability in the marked clique may not be the best approach in general. There may exist graphs where maximizing the probability in the marked clique is significantly worse than maximizing the probability at some subset of marked clique. For the weighted barbell graph with $w_-$ as its first stage, at least, there is no useful two-stage algorithm, and this is summarized in the bottom half of the last column of \tref{table:summary}.


\section{\label{sec:conclusion}Conclusion}

In this paper, we showed that a single-excitation spin network in the Heisenberg model can effect a generalized Laplacian quantum walk on a signed weighted graph, where the generalized Laplacian equals the standard Laplacian plus a real-valued multiple of the degree matrix. This generalized Laplacian encompasses the standard Laplacian, adjacency matrix, and signless Laplacian as special cases. We investigated how this generalized Laplacian quantum walk searches a weighted barbell graph, with an external magnetic field implementing an oracle, and we derived a critical jumping rate for the algorithm. We showed that when the weight of the bridge takes a noncritical value, the search algorithm reaches a success probability of $1/2$. We also showed that two critical weights $w_\pm$ generally exist (except for the standard and signless Laplacians, which have one), which could be positive or negative, where the success probability is boosted to $0.820$ and $0.843$. By using these critical weights to maximize the probability in certain vertices, and then using a noncritical weight to focus the probability onto the marked vertex, the $w_+$ algorithm can further improve the success probability to $0.996$, whereas the $w_-$ algorithm remains at $0.843$. All of the algorithms search in $O(\sqrt{N})$ time, with different constant factors and success probabilities. These results demonstrate that regardless of the multiple of the degree matrix used in the generalized Laplacian, the weight of bridge in the barbell graph can be chosen judiciously to improve the success probability over the unweighted case. If one wants the greatest success probability in a single run of the algorithm, then $0.996$ can be obtained by using $w_+$ followed by $w = 1$ in a two-stage algorithm, but if one cannot dynamically change $w$ during the evolution, then using $w_-$ for a single-stage algorithm and achieving a success probability of $0.843$ is preferred.


\begin{acknowledgments}
	This material is based upon work supported in part by the National Science Foundation EPSCoR Cooperative Agreement OIA-2044049, Nebraska’s EQUATE collaboration. Any opinions, findings, and conclusions or recommendations expressed in this material are those of the author(s) and do not necessarily reflect the views of the National Science Foundation.
\end{acknowledgments}


\appendix

\section{\label{appendix:wplus-twostage-ab}Two-stage algorithm when \texorpdfstring{$w = w_+$}{w = w+} maximizing the probability in \texorpdfstring{$a$}{a} and \texorpdfstring{$b$}{b}}

In this appendix, we prove that the two-stage algorithm in \sref{subsec:wplus}, which was obtained from \cite{Wong41} by maximizing the probability in the marked clique, can also be obtained from maximizing the probability in vertices $a$ and $b$ alone, i.e., without consideration of the probability in vertex $c$. From \cite{Wong41}, when $w = w_+$, the probability in each type of vertex at time $t$ is given by their (23), and adding the probability at vertex $a$ with the probability at vertex $b$, we have
\begin{align*}
    p_{ab}(t)
        &= \frac{1}{8} \left\{ \left[ \sqrt{2+\sqrt{2}} \sin \left( \sqrt{\frac{2+\sqrt{2}}{N}} t \right) \right. \right. \\
        &\qquad\qquad- \left. \sqrt{2-\sqrt{2}} \sin \left( \sqrt{\frac{2-\sqrt{2}}{N}} t \right) \right]^2 \\
        &\qquad+ \left[ \left( 1 + \sqrt{2} \right) \cos \left( \sqrt{\frac{2+\sqrt{2}}{N}} t \right) \right. \\
        &\qquad\qquad+ \left. \left. \left( 1 - \sqrt{2} \right) \cos \left( \sqrt{\frac{2-\sqrt{2}}{N}} t \right) \right]^2 \right\}.
\end{align*}
The derivative of this simplifies to
\begin{align*}
    \frac{dp_{ab}}{dt}
        &= \frac{-1}{4\sqrt{N}} \left[ \sqrt{2+\sqrt{2}} \sin \left( \sqrt{\frac{2+\sqrt{2}}{N}} t \right) \right. \\
        &\qquad\qquad+ \left. \sqrt{2-\sqrt{2}} \sin \left( \sqrt{\frac{2-\sqrt{2}}{N}} t \right) \right] \\
        &\qquad\enspace \times \left[ \left( 1 + \sqrt{2} \right) \cos \left( \sqrt{\frac{2+\sqrt{2}}{N}} t \right) \right. \\
        &\qquad\qquad+ \left. \left( 1 - \sqrt{2} \right) \cos \left( \sqrt{\frac{2-\sqrt{2}}{N}} t \right) \right].
\end{align*}
Setting this equal to zero to find the maximum probability in the $a$ and $b$ vertices, the relevant zero comes from the sine portion, i.e.,
\begin{align*}
    &\sqrt{2+\sqrt{2}} \sin \left( \sqrt{\frac{2+\sqrt{2}}{N}} t \right) \\
    &\quad+ \sqrt{2-\sqrt{2}} \sin \left( \sqrt{\frac{2-\sqrt{2}}{N}} t \right) = 0.
\end{align*}
This is the same transcendental equation from \cite{Wong41}, above their (26), that was obtained by maximizing the probability in the marked clique, and so it has the same solution, i.e., the first stage runs for time $t = 3.265 \sqrt{N}$. So, the two-stage algorithm is the same whether the first stage maximizes the probability in the marked clique or the probability in vertices $a$ and $b$, as reported in the main text.


\section{\label{appendix:wminus-twostage-abc}Two-stage algorithm when \texorpdfstring{$w = w_-$}{w = w-} maximizing the probability in the marked clique}

In this appendix, we prove the behavior of the two-stage algorithm with $w = w_-$ and $w = 1$, with the transition between them determined by the time at which the probability in the marked clique is maximized. During the first stage, for large $N$, the probability in the marked clique can be obtained by adding $p_a$, $p_b$, and $p_c$ from \eqref{eq:probs}, i.e.,
\[ p_{abc}(t) = p_a(t) + p_b(t) + p_c(t). \]
Taking the derivative of this and setting it equal to zero, we numerically find that $p_{abc}$ reaches a maximum at time
\begin{equation}
    \label{eq:wminus-twostage-abc-t1}
    t_1 = 6.097 \sqrt{N},
\end{equation}
with a maximum probability of 
\[ p_{abc}(t_1) = 0.917. \]
This is consistent with \fref{fig:prob-time-N1200-4-w300-types}, where around time $6.097 \sqrt{1200} = 211.2$, the probability in the marked clique reaches $0.917$. The state of the system at the end of the first stage is $\ket{\psi(t_1)}$ and can be obtained from \eqref{eq:state-at-time-t}, and we drop the overall phase of $e^{it}$ to obtain
\begin{align*}
    \ket{\psi(t_1)} 
        &= -0.913 i \ket{a} + 0.0782 \ket{b} + 0.277 i \ket{c} \\
        &\quad - 0.277 i \ket{d} - 0.0782 \ket{e}.
\end{align*}
Note the success probability at this time is
\[ p_a(t_1) = 0.834, \]
and the probability in the $a$ and $b$ vertices at this time is
\[ p_a(t_1) + p_b(t_1) = 0.840. \]
We will prove next that this probability of $0.840$ all collects at the marked vertex during the second stage.

For the second stage of the algorithm, we choose a noncritical weight, such as $w = 1$, where the asymptotic eigenvectors and eigenvalues of the search Hamiltonian are \eqref{eq:eigensystem-noncritical}. Expressing $\ket{\psi(t_1)}$ as a linear combination of these eigenvectors,
\begin{align*}
    \ket{\psi(t_1)}
        &= -0.648e^{1.656i} \ket{\psi_0} - 0.0782 \ket{\psi_1} \\ 
        &\quad + 0.648e^{-1.656i} \ket{\psi_2} + 0.392i \ket{\psi_4}.
\end{align*}
Then the state at time $t > t_1$ is given by evolving $\ket{\psi_1}$ for time $\Delta t = t - t_1$:
\begin{widetext}
\begin{align*}
    \ket{\psi(t)} 
    &= e^{-iH\Delta t} \ket{\psi(t_1)} \displaybreak[0] \\
    &= -0.648e^{1.656i} e^{-iE_0\Delta t} \ket{\psi_0} - 0.0782 e^{-i E_1 \Delta t} \ket{\psi_1} + 0.648e^{-1.656i} e^{-i E_2 \Delta t} \ket{\psi_2} + 0.392i e^{-i E_4 \Delta t} \ket{\psi_4} \displaybreak[0] \\
    &= -0.648e^{1.656i} e^{-i(-1-\sqrt{2/N})\Delta t} \frac{1}{\sqrt{2}} (\ket{a}-\ket{b}) - 0.0782 e^{i \Delta t} \ket{e} \\
        &\quad + 0.648e^{-1.656i} e^{-i(-1+\sqrt{2/N})\Delta t} \frac{1}{\sqrt{2}} (\ket{a}+\ket{b}) + 0.392i e^{2i(\alpha-2)w \Delta t / N} \frac{1}{\sqrt{2}} \left( \ket{c} - \ket{d} \right) \displaybreak[0] \\
    &= \frac{0.648}{\sqrt{2}} e^{i \Delta t} \left[ -e^{i(1.656+\sqrt{2/N}\Delta t)} + e^{i(-1.656-\sqrt{2/N}\Delta t)} \right] \ket{a} \\
        &\quad + \frac{0.648}{\sqrt{2}} e^{i \Delta t} \left[ e^{i(1.656+\sqrt{2/N}\Delta t)} + e^{i(-1.656-\sqrt{2/N}\Delta t)} \right] \ket{b} + \frac{0.392i}{\sqrt{2}} e^{2i(\alpha-2)w \Delta t / N} \ket{c} \\
        &\quad - \frac{0.392i}{\sqrt{2}} e^{2i(\alpha-2)w \Delta t / N} \ket{d} - 0.0782 e^{i \Delta t} \ket{e} \displaybreak[0] \\
    &= -\frac{0.648}{\sqrt{2}} e^{i \Delta t} 2i \sin \left( \sqrt{\frac{2}{N}}\Delta t + 1.656 \right) \ket{a} + \frac{0.648}{\sqrt{2}} e^{i \Delta t} 2 \cos \left (\sqrt{\frac{2}{N}}\Delta t + 1.656 \right) \ket{b} \\
        &\quad+ \frac{0.392i}{\sqrt{2}} e^{2i(\alpha-2)w \Delta t / N} \ket{c} - \frac{0.392i}{\sqrt{2}} e^{2i(\alpha-2)w \Delta t / N} \ket{d} - 0.0782 e^{i \Delta t} \ket{e}.
\end{align*}
\end{widetext}
Then, taking the norm-square of each amplitude, the probability at each type of vertex at time $t = t_1 + \Delta t$ is
\begin{align*}
    p_a(t) &= 0.840 \sin^2 \left( \sqrt{\frac{2}{N}}\Delta t + 1.656 \right), \displaybreak[0] \\
    p_b(t) &= 0.840 \cos^2 \left( \sqrt{\frac{2}{N}}\Delta t + 1.656 \right), \displaybreak[0] \\
    p_c(t) &= 0.0767, \displaybreak[0] \\
    p_d(t) &= 0.0767, \displaybreak[0] \\
    p_e(t) &= 0.00612.
\end{align*}
Since $p_a(t)$ is a squared sinusoidal function with amplitude $0.840$, during the second stage of the algorithm, the success probability reaches a maximum of $0.840$, as reported in the main text. The time that this takes is the value of $\Delta t$, call it $t_2$, that makes the argument of the sine equal to $3\pi/2$, i.e.,
\[ \sqrt{\frac{2}{N}} t_2 + 1.656 = \frac{3\pi}{2}. \]
The solution to this is
\[ t_2 = 2.161 \sqrt{N}. \]
Combining this with $t_1 = 6.097 \sqrt{N}$ \eqref{eq:wminus-twostage-abc-t1}, the total runtime of the two-stage algorithm is
\[ t_1 + t_2 = 7.753 \sqrt{N}. \]


\section{\label{appendix:wminus-twostage-ab}Two-stage algorithm when \texorpdfstring{$w = w_-$}{w = w-} maximizing the probability in \texorpdfstring{$a$}{a} and \texorpdfstring{$b$}{b}}

In this appendix, we prove the behavior of the two-stage algorithm with $w = w_-$ and $w = 1$, with the transition between them determined by the time at which the probability in $a$ and $b$ is maximized. We look at the probability in $a$ and $b$ from \eqref{eq:probs}, i.e.,
\[ p_{ab}(t) = p_a(t) + p_b(t). \]
Taking the derivative and setting it equal to zero, we see that $p_{ab}(t)$ is maximized at the same time that $p_a(t)$ is maximized, i.e., at time $t_1 = t_*$ from \eqref{eq:runtime}. From \eqref{eq:probs}, $p_b(t_*) = 0$, and so $p_{ab}(t_*) = p_a(t_*)$. That is, the time that we transition between the two stages is the same as the runtime of the single-stage algorithm. Using \eqref{eq:state-at-time-t} and dropping the global phase of $e^{it}$, the state of the system at this transition time is
\[ \ket{\psi(t_1)} = -0.918 i \ket{a} + 0.269 i \ket{c} - 0.269 i \ket{d} - 0.111 \ket{e}. \]

For the second stage of the algorithm, we choose a noncritical weight, such as $w = 1$, where the asymptotic eigenvectors and eigenvalues of the search algorithm are \eqref{eq:eigensystem-noncritical}. Expressing $\ket{\psi(t_*)}$ as a linear combination of these eigenvectors,
\begin{align*}
    \ket{\psi(t_*)}
        &= -0.649 i \ket{\psi_0} - 0.111 \ket{\psi_1} \\
        &\quad - 0.649 i \ket{\psi_2} + 0.380 i \ket{\psi_4}.
\end{align*}
Then, the state at time $t = t_1 + \Delta t$ is
\begin{align*}
    \ket{\psi(t)} 
    &= e^{-iH\Delta t} \ket{\psi(t_1)} \displaybreak[0] \\
    &= -0.649i e^{-iE_0\Delta t} \ket{\psi_0} - 0.111 e^{-i E_1 \Delta t} \ket{\psi_1} \\
        &\quad - 0.649i e^{-i E_2 \Delta t} \ket{\psi_2} + 0.380 i e^{-iE_4 \Delta t} \ket{\psi_4} \displaybreak[0] \\
    &= -0.649i e^{-i(-1-\sqrt{2/N})\Delta t} \frac{1}{\sqrt{2}} (\ket{a}+\ket{b}) \\
        &\quad - 0.111 e^{i \Delta t} \ket{e} \\
        &\quad - 0.649i e^{-i(-1+\sqrt{2/N})\Delta t} \frac{1}{\sqrt{2}} (\ket{a}-\ket{b}) \\
        &\quad + 0.380i e^{2i(\alpha-2)w \Delta t / N} \frac{1}{\sqrt{2}} \left( \ket{c} - \ket{d} \right) \displaybreak[0] \\
    &= \frac{-0.649i}{\sqrt{2}} e^{i\Delta t} \left( e^{i\sqrt{2/N}\Delta t} + e^{-i\sqrt{2/N}\Delta t}\right) \ket{a} \\
        &\quad + \frac{0.649i}{\sqrt{2}} e^{i\Delta t} \left( -e^{i\sqrt{2/N}\Delta t} + e^{-i\sqrt{2/N}\Delta t} \right) \ket{b} \\
        &\quad + \frac{0.380i}{\sqrt{2}} e^{2i(\alpha-2)w \Delta t / N} \ket{c} \\
        &\quad - \frac{0.380i}{\sqrt{2}} e^{2i(\alpha-2)w \Delta t / N} \ket{d} \\
        &\quad - 0.111 e^{i\Delta t} \ket{e} \displaybreak[0] \\
    &= \frac{-0.649i}{\sqrt{2}} e^{i\Delta t} 2 \cos \left( \sqrt{\frac{2}{N}} \Delta t \right) \ket{a} \\
        &\quad - \frac{0.649i}{\sqrt{2}} e^{i\Delta t} 2i \sin \left( \sqrt{\frac{2}{N}} \Delta t \right) \ket{b} \\
        &\quad + \frac{0.380i}{\sqrt{2}} e^{2i(\alpha-2)w \Delta t / N} \ket{c} \\
        &\quad - \frac{0.380i}{\sqrt{2}} e^{2i(\alpha-2)w \Delta t / N} \ket{d} \\
        &\quad - 0.111 e^{i\Delta t} \ket{e}.
\end{align*}
Taking the norm squared of each amplitude, the probability at each type of vertex is
\begin{align*}
    p_a(\Delta t) &= 0.843 \cos \left( \sqrt{\frac{2}{N}} \Delta t \right)^2, \displaybreak[0] \\
    p_b(\Delta t) &= 0.843 \sin \left( \sqrt{\frac{2}{N}} \Delta t \right)^2, \displaybreak[0] \\
    p_c(\Delta t) &= 0.0723,\displaybreak[0] \\
    p_d(\Delta t) &= 0.0723, \displaybreak[0] \\
    p_e(\Delta t) &= 0.0124.
\end{align*}
Since $p_a(t)$ is a squared sinusoidal function with amplitude $0.843$, during the second stage of the algorithm, the success probability reaches a maximum of $0.843$. Since the transition between the stages occurs at this maximum, one can stop after the first stage and skip the second stage, or run the second stage for a period of $t_2 = \pi\sqrt{N/2}$. Combining this with the first stage's runtime of $t_1 = t_* \approx 6.011 \sqrt{N}$, the total runtime of the two-stage algorithm is
\[ t_1 + t_2 = 7.582 \sqrt{N}. \]


\bibliography{refs}

\begin{thebibliography}{14}%
\makeatletter
\providecommand \@ifxundefined [1]{%
 \@ifx{#1\undefined}
}%
\providecommand \@ifnum [1]{%
 \ifnum #1\expandafter \@firstoftwo
 \else \expandafter \@secondoftwo
 \fi
}%
\providecommand \@ifx [1]{%
 \ifx #1\expandafter \@firstoftwo
 \else \expandafter \@secondoftwo
 \fi
}%
\providecommand \natexlab [1]{#1}%
\providecommand \enquote  [1]{``#1''}%
\providecommand \bibnamefont  [1]{#1}%
\providecommand \bibfnamefont [1]{#1}%
\providecommand \citenamefont [1]{#1}%
\providecommand \href@noop [0]{\@secondoftwo}%
\providecommand \href [0]{\begingroup \@sanitize@url \@href}%
\providecommand \@href[1]{\@@startlink{#1}\@@href}%
\providecommand \@@href[1]{\endgroup#1\@@endlink}%
\providecommand \@sanitize@url [0]{\catcode `\\12\catcode `\$12\catcode `\&12\catcode `\#12\catcode `\^12\catcode `\_12\catcode `\%12\relax}%
\providecommand \@@startlink[1]{}%
\providecommand \@@endlink[0]{}%
\providecommand \url  [0]{\begingroup\@sanitize@url \@url }%
\providecommand \@url [1]{\endgroup\@href {#1}{\urlprefix }}%
\providecommand \urlprefix  [0]{URL }%
\providecommand \Eprint [0]{\href }%
\providecommand \doibase [0]{https://doi.org/}%
\providecommand \selectlanguage [0]{\@gobble}%
\providecommand \bibinfo  [0]{\@secondoftwo}%
\providecommand \bibfield  [0]{\@secondoftwo}%
\providecommand \translation [1]{[#1]}%
\providecommand \BibitemOpen [0]{}%
\providecommand \bibitemStop [0]{}%
\providecommand \bibitemNoStop [0]{.\EOS\space}%
\providecommand \EOS [0]{\spacefactor3000\relax}%
\providecommand \BibitemShut  [1]{\csname bibitem#1\endcsname}%
\let\auto@bib@innerbib\@empty
\bibitem [{\citenamefont {Farhi}\ and\ \citenamefont {Gutmann}(1998)}]{FG1998b}%
  \BibitemOpen
  \bibfield  {author} {\bibinfo {author} {\bibfnamefont {E.}~\bibnamefont {Farhi}}\ and\ \bibinfo {author} {\bibfnamefont {S.}~\bibnamefont {Gutmann}},\ }\bibfield  {title} {\bibinfo {title} {Quantum computation and decision trees},\ }\href {https://doi.org/10.1103/PhysRevA.58.915} {\bibfield  {journal} {\bibinfo  {journal} {Phys. Rev. A}\ }\textbf {\bibinfo {volume} {58}},\ \bibinfo {pages} {915} (\bibinfo {year} {1998})}\BibitemShut {NoStop}%
\bibitem [{\citenamefont {Childs}(2009)}]{Childs2009}%
  \BibitemOpen
  \bibfield  {author} {\bibinfo {author} {\bibfnamefont {A.~M.}\ \bibnamefont {Childs}},\ }\bibfield  {title} {\bibinfo {title} {Universal computation by quantum walk},\ }\href {https://doi.org/10.1103/PhysRevLett.102.180501} {\bibfield  {journal} {\bibinfo  {journal} {Phys. Rev. Lett.}\ }\textbf {\bibinfo {volume} {102}},\ \bibinfo {pages} {180501} (\bibinfo {year} {2009})}\BibitemShut {NoStop}%
\bibitem [{\citenamefont {Childs}\ and\ \citenamefont {Goldstone}(2004)}]{CG2004}%
  \BibitemOpen
  \bibfield  {author} {\bibinfo {author} {\bibfnamefont {A.~M.}\ \bibnamefont {Childs}}\ and\ \bibinfo {author} {\bibfnamefont {J.}~\bibnamefont {Goldstone}},\ }\bibfield  {title} {\bibinfo {title} {Spatial search by quantum walk},\ }\href {https://doi.org/10.1103/PhysRevA.70.022314} {\bibfield  {journal} {\bibinfo  {journal} {Phys. Rev. A}\ }\textbf {\bibinfo {volume} {70}},\ \bibinfo {pages} {022314} (\bibinfo {year} {2004})}\BibitemShut {NoStop}%
\bibitem [{\citenamefont {Alvir}\ \emph {et~al.}(2016)\citenamefont {Alvir}, \citenamefont {Dever}, \citenamefont {Lovitz}, \citenamefont {Myer}, \citenamefont {Tamon}, \citenamefont {Xu},\ and\ \citenamefont {Zhan}}]{Alvir2016}%
  \BibitemOpen
  \bibfield  {author} {\bibinfo {author} {\bibfnamefont {R.}~\bibnamefont {Alvir}}, \bibinfo {author} {\bibfnamefont {S.}~\bibnamefont {Dever}}, \bibinfo {author} {\bibfnamefont {B.}~\bibnamefont {Lovitz}}, \bibinfo {author} {\bibfnamefont {J.}~\bibnamefont {Myer}}, \bibinfo {author} {\bibfnamefont {C.}~\bibnamefont {Tamon}}, \bibinfo {author} {\bibfnamefont {Y.}~\bibnamefont {Xu}},\ and\ \bibinfo {author} {\bibfnamefont {H.}~\bibnamefont {Zhan}},\ }\bibfield  {title} {\bibinfo {title} {Perfect state transfer in laplacian quantum walk},\ }\href {https://doi.org/10.1007/s10801-015-0642-x} {\bibfield  {journal} {\bibinfo  {journal} {Journal of Algebraic Combinatorics}\ }\textbf {\bibinfo {volume} {43}},\ \bibinfo {pages} {801} (\bibinfo {year} {2016})}\BibitemShut {NoStop}%
\bibitem [{\citenamefont {Novo}\ \emph {et~al.}(2015)\citenamefont {Novo}, \citenamefont {Chakraborty}, \citenamefont {Mohseni}, \citenamefont {Neven},\ and\ \citenamefont {Omar}}]{Novo2015}%
  \BibitemOpen
  \bibfield  {author} {\bibinfo {author} {\bibfnamefont {L.}~\bibnamefont {Novo}}, \bibinfo {author} {\bibfnamefont {S.}~\bibnamefont {Chakraborty}}, \bibinfo {author} {\bibfnamefont {M.}~\bibnamefont {Mohseni}}, \bibinfo {author} {\bibfnamefont {H.}~\bibnamefont {Neven}},\ and\ \bibinfo {author} {\bibfnamefont {Y.}~\bibnamefont {Omar}},\ }\bibfield  {title} {\bibinfo {title} {Systematic dimensionality reduction for quantum walks: Optimal spatial search and transport on non-regular graphs},\ }\href {https://doi.org/10.1038/srep13304} {\bibfield  {journal} {\bibinfo  {journal} {Sci. Rep.}\ }\textbf {\bibinfo {volume} {5}},\ \bibinfo {pages} {13304} (\bibinfo {year} {2015})}\BibitemShut {NoStop}%
\bibitem [{\citenamefont {Godsil}(2012)}]{Godsil2012}%
  \BibitemOpen
  \bibfield  {author} {\bibinfo {author} {\bibfnamefont {C.}~\bibnamefont {Godsil}},\ }\bibfield  {title} {\bibinfo {title} {State transfer on graphs},\ }\href {https://doi.org/https://doi.org/10.1016/j.disc.2011.06.032} {\bibfield  {journal} {\bibinfo  {journal} {Discrete Mathematics}\ }\textbf {\bibinfo {volume} {312}},\ \bibinfo {pages} {129 } (\bibinfo {year} {2012})},\ \bibinfo {note} {{A}lgebraic Graph Theory — A Volume Dedicated to Gert Sabidussi on the Occasion of His 80th Birthday}\BibitemShut {NoStop}%
\bibitem [{\citenamefont {Farhi}\ \emph {et~al.}(2008)\citenamefont {Farhi}, \citenamefont {Goldstone},\ and\ \citenamefont {Gutmann}}]{FGG2008}%
  \BibitemOpen
  \bibfield  {author} {\bibinfo {author} {\bibfnamefont {E.}~\bibnamefont {Farhi}}, \bibinfo {author} {\bibfnamefont {J.}~\bibnamefont {Goldstone}},\ and\ \bibinfo {author} {\bibfnamefont {S.}~\bibnamefont {Gutmann}},\ }\bibfield  {title} {\bibinfo {title} {A quantum algorithm for the {H}amiltonian {NAND} tree},\ }\href {https://doi.org/10.4086/toc.2008.v004a008} {\bibfield  {journal} {\bibinfo  {journal} {Theory Comput.}\ }\textbf {\bibinfo {volume} {4}},\ \bibinfo {pages} {169} (\bibinfo {year} {2008})}\BibitemShut {NoStop}%
\bibitem [{\citenamefont {Childs}\ \emph {et~al.}(2003)\citenamefont {Childs}, \citenamefont {Cleve}, \citenamefont {Deotto}, \citenamefont {Farhi}, \citenamefont {Gutmann},\ and\ \citenamefont {Spielman}}]{Childs2003}%
  \BibitemOpen
  \bibfield  {author} {\bibinfo {author} {\bibfnamefont {A.~M.}\ \bibnamefont {Childs}}, \bibinfo {author} {\bibfnamefont {R.}~\bibnamefont {Cleve}}, \bibinfo {author} {\bibfnamefont {E.}~\bibnamefont {Deotto}}, \bibinfo {author} {\bibfnamefont {E.}~\bibnamefont {Farhi}}, \bibinfo {author} {\bibfnamefont {S.}~\bibnamefont {Gutmann}},\ and\ \bibinfo {author} {\bibfnamefont {D.~A.}\ \bibnamefont {Spielman}},\ }\bibfield  {title} {\bibinfo {title} {Exponential algorithmic speedup by a quantum walk},\ }in\ \href {https://doi.org/10.1145/780542.780552} {\emph {\bibinfo {booktitle} {Proceedings of the 35th Annual ACM Symposium on Theory of Computing}}},\ \bibinfo {series and number} {STOC '03}\ (\bibinfo  {publisher} {ACM},\ \bibinfo {address} {New York, NY, USA},\ \bibinfo {year} {2003})\ pp.\ \bibinfo {pages} {59--68}\BibitemShut {NoStop}%
\bibitem [{\citenamefont {McLaughlin}\ and\ \citenamefont {Wong}(2025)}]{Wong43}%
  \BibitemOpen
  \bibfield  {author} {\bibinfo {author} {\bibfnamefont {M.~E.}\ \bibnamefont {McLaughlin}}\ and\ \bibinfo {author} {\bibfnamefont {T.~G.}\ \bibnamefont {Wong}},\ }\bibfield  {title} {\bibinfo {title} {Quantum search with the signless {L}aplacian},\ }\href {https://doi.org/10.1103/PhysRevA.111.032430} {\bibfield  {journal} {\bibinfo  {journal} {Phys. Rev. A}\ }\textbf {\bibinfo {volume} {111}},\ \bibinfo {pages} {032430} (\bibinfo {year} {2025})}\BibitemShut {NoStop}%
\bibitem [{\citenamefont {Bose}\ \emph {et~al.}(2009)\citenamefont {Bose}, \citenamefont {Casaccino}, \citenamefont {Mancini},\ and\ \citenamefont {Severini}}]{Bose2009}%
  \BibitemOpen
  \bibfield  {author} {\bibinfo {author} {\bibfnamefont {S.}~\bibnamefont {Bose}}, \bibinfo {author} {\bibfnamefont {A.}~\bibnamefont {Casaccino}}, \bibinfo {author} {\bibfnamefont {S.}~\bibnamefont {Mancini}},\ and\ \bibinfo {author} {\bibfnamefont {S.}~\bibnamefont {Severini}},\ }\bibfield  {title} {\bibinfo {title} {Communication in {XYZ} all-to-all quantum networks with a missing link},\ }\href {https://doi.org/10.1142/S0219749909005389} {\bibfield  {journal} {\bibinfo  {journal} {Int. J. Quantum Inf.}\ }\textbf {\bibinfo {volume} {07}},\ \bibinfo {pages} {713} (\bibinfo {year} {2009})}\BibitemShut {NoStop}%
\bibitem [{\citenamefont {Meyer}\ and\ \citenamefont {Wong}(2015)}]{Wong7}%
  \BibitemOpen
  \bibfield  {author} {\bibinfo {author} {\bibfnamefont {D.~A.}\ \bibnamefont {Meyer}}\ and\ \bibinfo {author} {\bibfnamefont {T.~G.}\ \bibnamefont {Wong}},\ }\bibfield  {title} {\bibinfo {title} {Connectivity is a poor indicator of fast quantum search},\ }\href {https://doi.org/10.1103/PhysRevLett.114.110503} {\bibfield  {journal} {\bibinfo  {journal} {Phys. Rev. Lett.}\ }\textbf {\bibinfo {volume} {114}},\ \bibinfo {pages} {110503} (\bibinfo {year} {2015})}\BibitemShut {NoStop}%
\bibitem [{\citenamefont {Duda}\ and\ \citenamefont {Wong}(2024)}]{Wong41}%
  \BibitemOpen
  \bibfield  {author} {\bibinfo {author} {\bibfnamefont {J.}~\bibnamefont {Duda}}\ and\ \bibinfo {author} {\bibfnamefont {T.~G.}\ \bibnamefont {Wong}},\ }\bibfield  {title} {\bibinfo {title} {Searching weighted barbell graphs with {L}aplacian and adjacency quantum walks},\ }\href {https://doi.org/10.1103/PhysRevA.110.042417} {\bibfield  {journal} {\bibinfo  {journal} {Phys. Rev. A}\ }\textbf {\bibinfo {volume} {110}},\ \bibinfo {pages} {042417} (\bibinfo {year} {2024})}\BibitemShut {NoStop}%
\bibitem [{\citenamefont {Heisenberg}(1928)}]{Heisenberg1928}%
  \BibitemOpen
  \bibfield  {author} {\bibinfo {author} {\bibfnamefont {W.}~\bibnamefont {Heisenberg}},\ }\bibfield  {title} {\bibinfo {title} {Zur theorie des ferromagnetismus},\ }\href {https://doi.org/10.1007/BF01328601} {\bibfield  {journal} {\bibinfo  {journal} {Zeitschrift f{\"u}r Physik}\ }\textbf {\bibinfo {volume} {49}},\ \bibinfo {pages} {619} (\bibinfo {year} {1928})}\BibitemShut {NoStop}%
\bibitem [{\citenamefont {Baxter}(1982)}]{Baxter1982}%
  \BibitemOpen
  \bibfield  {author} {\bibinfo {author} {\bibfnamefont {R.~J.}\ \bibnamefont {Baxter}},\ }\href@noop {} {\emph {\bibinfo {title} {{Exactly Solved Models in Statistical Mechanics}}}}\ (\bibinfo  {publisher} {Academic Press},\ \bibinfo {address} {London},\ \bibinfo {year} {1982})\BibitemShut {NoStop}%
\end{thebibliography}%

\end{document}